\documentclass[a4paper, 11pt]{article}
\usepackage{amsfonts,slashed}
\usepackage{float}
\usepackage{cite}
\usepackage{latexsym}
\usepackage{amsfonts}
\usepackage{amsmath}
\usepackage{epsfig}
\usepackage{amssymb}
\usepackage{amsthm}
\usepackage{xcolor}
\usepackage{hyperref}
\definecolor{darkorange}{RGB}{166, 88, 0}
\hypersetup{colorlinks=true,
        pdfstartview=FitV,
        linkcolor= darkorange,
        citecolor= darkorange, 
        urlcolor= gray!60!black,
        hypertexnames=false,
        linktoc=page}
\usepackage{graphicx,tikz}
\usetikzlibrary{arrows.meta,arrows}
\usepackage{multirow}
\usepackage{textcomp}
\usepackage{mathtools}
\usepackage{stmaryrd}
\usepackage{enumerate}
\usepackage[shortlabels]{enumitem}

\usepackage{xspace}
\usepackage{mathrsfs}
\usepackage{nicefrac}

\DeclareFontFamily{OMS}{rsfs}{\skewchar\font'60}
\DeclareFontShape{OMS}{rsfs}{m}{n}{<-5>rsfs5 <5-7>rsfs7 <7->rsfs10 }{}
\DeclareSymbolFont{rsfs}{OMS}{rsfs}{m}{n}
\DeclareSymbolFontAlphabet{\Scr}{rsfs}

\numberwithin{equation}{section}
\def\be{\begin{equation}}
	\def\ee{\end{equation}}
\def\ba{\begin{array}}
	\def\ea{\end{array}}

\newcommand{\bea}{\begin{eqnarray}}
	\newcommand{\eea}{\end{eqnarray}}

\newcommand{\ts}[1]{\widetilde{#1}}

\newcommand{\ff}[1]{\mathfrak{#1}}

\newcommand{\ov}[1]{\overline{#1}}

\newcommand{\SU}[1]{\mathrm{SU}( #1 )}
\newcommand{\SL}[1]{\mathrm{SL}( #1 )}
\newcommand{\GL}[1]{\mathrm{GL}( #1 )}
\newcommand{\SO}[1]{\mathrm{SO}( #1 )}

\newcommand{\U}[1]{\mathrm{U}(#1)}

\newcommand{\EE}{\ensuremath{E_{8(8)}}\xspace}

\newcommand{\En}[1]{E_{#1(#1)}}

\newcommand{\mbf}[1]{\mathbf{#1}}

\newcommand{\gL}{\mathcal{L}}
\newcommand{\gM}{\mathcal{M}}

\newcommand{\cU}{{\cal U}}
\newcommand{\UI}{\left(U^{-1}\right)}

\newcommand{\fl}[1]{\ov{#1}}

\newcommand{\cA}{\mathcal{A}}
\newcommand{\cB}{\mathcal{B}}
\newcommand{\cC}{\mathcal{C}}
\newcommand{\dg}{\vert g \vert}

\newcommand{\RO}{\mathbb{R}^+_{\SL{4}_1}}
\newcommand{\RT}{\mathbb{R}^+_{\SL{4}_2}}

\newcommand{\M}{\ensuremath{\mathcal{M}}\xspace}
\newcommand{\A}{\ensuremath{\mathcal{A}}\xspace}
\renewcommand{\d}{\ensuremath{\mathrm{d}}\xspace}

\definecolor{darkorange}{RGB}{166, 88, 0}

\begin{document}
	
\begin{titlepage}
	\vfill
	\begin{flushright}
		HU-EP-23/18
	\end{flushright}
	
	\vfill

	\begin{center}
		\baselineskip=16pt
		{\Large \bf Adding fluxes to consistent truncations: \\ IIB supergravity on ${\rm AdS}_3 \times S^3 \times S^3 \times S^1$}
		\vskip 2cm
		{\large \bf Camille Eloy$^a$\footnote{\tt camille.eloy@vub.be}, Michele Galli$^b$\footnote{\tt michele.galli@physik.hu-berlin.de},  Emanuel Malek$^b$\footnote{\tt emanuel.malek@physik.hu-berlin.de}}
		\vskip .6cm
		{\it $^a$ Theoretische Natuurkunde, Vrije Universiteit Brussel, and the Interational Solvay, \\
			Institutes, Pleinlaan 2, B-1050 Brussels, Belgium}
		\\ \ \\
		{\it $^b$ Institut f\"ur Physik, Humboldt-Universit\"at zu Berlin, \\IRIS Geb\"aude, Zum Gro\ss en Windkanal 2, 12489 Berlin, Germany \\ \ \\}
		\vskip 1cm
	\end{center}
	\vfill
		
	\begin{abstract}
		We use $\EE$ Exceptional Field Theory to construct the consistent truncation of IIB supergravity on $S^3 \times S^3 \times S^1$ to maximal 3-dimensional ${\cal N}=16$ gauged supergravity containing the ${\cal N}=(4,4)$ AdS$_3$ vacuum. We explain how to achieve this by adding a 7-form flux to the $S^1$ reduction of the dyonic $\En{7}$ truncation on $S^3 \times S^3$ previously constructed in the literature. Our truncation Ansatz includes, in addition to the ${\cal N}=(4,4)$ vacuum, a host of moduli breaking some or all of the supersymmetries. We explicitly construct the uplift of a subset of these to construct new supersymmetric and non-supersymmetric AdS$_3$ vacua of IIB string theory, which include a range of perturbatively stable non-supersymmetric 10-d vacua. Moreover, we show how the supersymmetric direction of the moduli space of AdS$_3$ vacua of six-dimensional gauged supergravity studied in \cite{Eloy:2021fhc} is compactified upon lifting to 10 dimensions, and find evidence of T-duality playing a role in global aspects of the moduli space. Along the way, we also derive the form of 3-dimensional ${\cal N}=16$ gauged supergravity in terms of the embedding tensor and rule out a 10-/11-dimensional origin of some 3-dimensional gauged supergravities.
	\end{abstract}
	\vskip 0cm
		
	\vfill
		
	\setcounter{footnote}{0} 
		
\end{titlepage}
	
\tableofcontents
	
\newpage

\section{Introduction}
Consistent truncations are a powerful technique that simplifies the dynamics of 10-/11-dimensional supergravity by focusing on a restricted subset of fields. The key principle behind a consistent truncation is to truncate to a subsector of fields, such that all solutions of the truncated theory correspond to solutions of the original 10-/11-dimensional supergravity theory. By considering only a, typically finite, subset of fields, consistent truncations provide a powerful tool to find complicated new 10-/11-dimensional supergravity solutions and to study their deformations. Consistent truncations have proven particularly powerful in the AdS/CFT correspondence, since all well-understood AdS vacua of string theory do not admit scale-separation and, thus, cannot be studied using the usual tools of lower-dimensional effective theories.

Exceptional Field Theory (ExFT) is a reformulation of 10-/11-dimensional supergravity that unifies the metric and flux degrees of freedom and thereby makes manifest an exceptional symmetry group\footnote{For the purposes of this paper, we do not draw a distinction between ExFT and Exceptional Generalised Geometry, since these agree when the ``section condition'' is solved, which we will always assume here.}. Over the last decade, this has proven extremely useful for constructing consistent truncations, leading to a number of new examples preserving various amounts of supersymmetry \cite{Hohm:2014qga,Lee:2014mla,Malek:2015hma,Baguet:2015sma,Baguet:2015iou,Lee:2015xga,Malek:2016bpu,Ciceri:2016dmd,Cassani:2016ncu,Inverso:2016eet,Malek:2017cle,Malek:2017njj,Malek:2018zcz,Malek:2019ucd,Cassani:2019vcl,Malek:2020jsa,Cassani:2020cod}. However, consistent truncations to three dimensions have, until recently \cite{Galli:2022idq}, remained largely unexplored. One reason is that the $\EE$ ExFT requires modifications for its local symmetry structure, \textit{i.e.} the generalised Lie derivative, to close into an algebra \cite{Hohm:2014fxa,Cederwall:2015ica}. Another is that there are no maximally supersymmetric AdS$_3$ vacua, \textit{i.e.} preserving 32 supercharges, of string theory, leaving no natural candidate for constructing a consistent truncation with vacua. Indeed, while the consistent truncations of 11-dimensional supergravity on $S^7$ and $S^4$ and of 10-dimensional supergravity on $S^5$ contain maximally supersymmetric AdS vacua and are captured by a universal Ansatz in ExFT \cite{Hohm:2014qga,Lee:2014mla}, the analogous $S^8$ truncation of 11-dimensional supergravity does not exist, while there are two $S^7$ truncations of 10-dimensional supergravity but neither contains a maximally symmetric vacuum\cite{Fischbacher:2003yw,Galli:2022idq}.

Nonetheless, there are half-maximal, \textit{i.e.} ${\cal N}=(4,4)$, supersymmetric AdS$_3$ vacua of string theory that are extremely intriguing. These are the AdS$_3 \times S^3 \times S^3 \times S^1$ and AdS$_3 \times S^3 \times T^4$ of IIB string theory (realising the ``large'' and ``small'' ${\cal N}=(4,4)$ superconformal symmetries), and, unlike in higher dimensions, can be supported by pure NS-NS flux. Not only does this mean that there are also heterotic versions of these vacua (preserving ${\cal N}=(4,0)$ supersymmetry), but also that they can be readily studied via the string worldsheet CFT \cite{Maldacena:2000hw,Eberhardt:2018ouy,Eberhardt:2019niq,Eberhardt:2019ywk}.

Here we will focus on the ${\cal N}=(4,4)$ AdS$_3 \times S^3 \times S^3 \times S^1$ vacuum of IIB string theory and show that it admits a consistent truncation to a 3-dimensional maximal gauged supergravity, which was first studied in \cite{Hohm:2005ui}. All vacua of this theory break at least half the supersymmetries, reflecting the absence of a maximally supersymmetric AdS$_3$ vacuum in string theory. Key to constructing this consistent truncation is to add new fluxes to the $S^1$ reduction of the consistent truncation of IIB supergravity on $S^3 \times S^3$ \cite{Inverso:2016eet}. We will also show that this procedure for adding flux fails to produce a consistent truncation for the $S^1$ reduction of other ``dyonic'' truncations to 4 dimensions \cite{Guarino:2015jca,Inverso:2016eet}.

Using our consistent truncation, we will show that the AdS$_3 \times S^3 \times S^3 \times S^1$ has a rich moduli space of symmetry- and supersymmetry-breaking deformations, including some that are analogous to the ``flat deformations'' studied for AdS$_4 \times S^5 \times S^1$ \cite{Guarino:2020gfe,Guarino:2021hrc,Giambrone:2021zvp,Giambrone:2021wsm}. While some of these deformations preserve some amounts of supersymmetry, others break all supersymmetries. Yet, surprisingly, at least a subset of vacua continue to be perturbatively stable within IIB supergravity, similar to \cite{Giambrone:2021wsm}. Our work also provides an uplift of the supersymmetric deformation of 6-dimensional supergravity on $S^3$ studied in \cite{Eloy:2021fhc} to IIB string theory. As we will show, some supersymmetric deformations that appear non-compact in 6-dimensional supergravity \cite{Eloy:2021fhc} are compactified within the full 10-dimensional supergravity theory. We will also find evidence of T-duality playing a role in the global properties of the moduli space.

Another technical result of our work is the derivation of the potential of 3-dimensional gauged supergravity \cite{Nicolai:2000sc,Nicolai:2001sv} in terms of the embedding tensor. The potential was previously only known in terms of fermion shift matrices. As a by-product, we find a constraint that must be obeyed by the embedding tensor of any 3-dimensional gauged supergravity that can be uplifted to 10-/11-dimensional supergravity. This allows to prove the lack of a higher-dimensional origin of some 3-dimensional gaugings, as we illustrate for an example.

The outline of our paper is as follows. We begin with a review of $\EE$ ExFT in section \ref{s:Review} and how to derive consistent truncations in this formalism in \ref{s:TruncationReview}. In section \ref{s:Potential} we derive the potential of 3-dimensional gauged supergravity in terms of the embedding tensor and prove the lack of higher-dimensional origin of some 3-dimensional theories. Then, we show how to add fluxes to consistent truncations in \ref{s:AddFlux}, allowing us to construct the consistent truncation of IIB supergravity on $S^3 \times S^3 \times S^1$. Using this truncation, we study the moduli space of the AdS$_3$ vacua in section \ref{s:Moduli}, before concluding in section \ref{s:Conclusions}. We end the paper by an example in appendix~\ref{app:rulingoutgauging} of how to rule out a higher-dimensional origin for a three-dimensional supergravity, a review of the construction of consistent truncations on $S^{3}$ in appendix~\ref{s:SLn} and a discussion of the $D^{1}(2,1;\alpha)$ supermultiplets in appendix~\ref{app:d21a}.

\section{Review of \texorpdfstring{\EE}{E8(8)} exceptional field theory} \label{s:Review}

The \EE exceptional field theory, first constructed in ref.~\cite{Hohm:2014fxa}, is an \EE duality-covariant formulation of type II and 11d supergravities. It is defined on a set of $3+248$ coordinates made of three-dimensional external coordinates $\{x^{\mu}\}$ and internal coordinates $\{Y^{M}\}$ in the 248-dimensional adjoint representation of \EE. The dependence of the fields on these coordinates is constrained by the ``section constraints''\footnote{We use the notation $\otimes$ to indicate that both derivatives may act on different functions.}
\begin{equation} \label{eq:sectionconstraints}
	\begin{cases}
		\eta^{MN}\partial_{M}\otimes\partial_{N} = 0\,,\\
		f^{MN}{}_{P}\,\partial_{M}\otimes\partial_{N} = 0\,,\\
		(\mathbb{P}_{3875})_{MN}{}^{KL}\partial_{K}\otimes\partial_{L} = 0\,,
	\end{cases}
\end{equation}
where $f_{MN}{}^{P}$ are the structure constants of \EE, $\eta_{MN} = \tfrac{1}{60}\,f_{MK}{}^{L}f_{NL}{}^{K}$ its Cartan-Killing metric and $(\mathbb{P}_{3875})_{MN}{}^{KL}$ is the projector on the representation $\mathbf{3875}$:
\begin{equation}
	(\mathbb{P}_{3875})_{MN}{}^{KL} = \frac{1}{7}\, \delta_{(M}{}^{K}\delta_{N)}{}^{L}-\frac{1}{56}\,\eta_{MN}\eta^{KL}-\frac{1}{14}\,f^{P}{}_{M}{}^{(K}f_{PN}{}^{L)}\,.
\end{equation}
Here and in the following, \EE indices are raised and lowered by the Cartan-Killing metric $\eta_{MN}$. The section constraints~\eqref{eq:sectionconstraints} ensure that the fields depends only on the 7- or 8-dimensional physical internal coordinates embedded in $Y^{M}$.

The theory describes the dynamics of the following bosonic fields:
\begin{equation}
	\left\{g_{\mu\nu}, \M_{MN}, A_{\mu}{}^{M}, B_{\mu\,M}\right\}\,,
\end{equation}
with $g_{\mu\nu}$ the 3-dimensional external metric, $\M_{MN}$ the generalized metric parametrizing the coset space $\EE/\SO{16}$ and the gauge fields $A_{\mu}{}^{M}$ and $B_{\mu\,M}$. It is a gauge theory, invariant under the generalized Lie derivative of parameters $\Upsilon=(\Lambda^{M},\Sigma_{M})$, whose action on a vector $V^{M}$ of weight $\lambda$ is given by
\begin{equation} \label{eq:genlie}
	{\cal L}_{\Upsilon} V^{M} = \Lambda^{N}\partial_{N}V^{M}-60\,(\mathbb{P}_{248})^{M}{}_{N}{}^{K}{}_{L}\,V^{N}\partial_{K}\Lambda^{L} + \lambda\, V^{M}\partial_{N}\Lambda^{N}+f^{MN}{}_{K}\Sigma_{N}V^{K}\,,
\end{equation}
with $(\mathbb{P}_{248})^{M}{}_{N}{}^{K}{}_{L}=(1/60)\,f^{M}{}_{NP}f^{PK}{}_{L}$ the projector on the adjoint representation. These transformations are well-defined (in particular, they close into an algebra) only if the parameters $\Sigma_{M}$ and the fields $B_{\mu\, M}$ are covariantly constrained: they have to satisfy algebraic constraints similar to eq.~\eqref{eq:sectionconstraints} and be compatibile with the partial derivatives. We require that
\begin{equation} \label{eq:sectionconstraint2}
	\begin{cases}
		\eta^{MN}C_{M}\otimes C'_{N} = 0\,,\\
		f^{MN}{}_{P}\,C_{M}\otimes C'_{N} = 0\,,\\
		(\mathbb{P}_{3875})_{MN}{}^{KL}C_{K}\otimes C'_{L} = 0\,,
	\end{cases}
	\forall\  C_{M}, C'_{M} \in \{\partial_{M}, \Sigma_{M}, B_{\mu\,M}\}\,.
\end{equation}

\paragraph{}
The bosonic action of \EE exceptional field theory is invariant under the transformations~\eqref{eq:genlie} and has the expression
\begin{equation} \label{eq:ExFTaction}
	S_{\rm ExFT} = \int d^{3}x\,d^{248}Y\,\sqrt{\vert g\vert} \bigg(\widehat{R}+\frac{1}{240}\,D_{\mu}\M_{MN}D^{\mu}\M^{MN}+\mathscr{L}_{\rm int}+\frac{1}{\sqrt{\vert g\vert}}\,\mathscr{L}_{\rm CS}\bigg)\,.
\end{equation}
$\widehat{R}$ is the \EE-covariantised Ricci scalar and the covariant derivative is defined as
\begin{equation}
	D_{\mu} = \partial_{\mu} - {\cal L}_{(A_{\mu}, B_{\mu})}\,.
\end{equation}
$\mathscr{L}_{\rm int}$ is a potential term depending only on internal derivatives, explicitly
\begin{equation} \label{eq:ExFTLint}
	\begin{aligned}
		\mathscr{L}_{\rm int} &= \frac{1}{240}\,\M^{MN}\partial_{M}\M^{KL}\partial_{N}\M_{KL} - \frac{1}{2}\,\M^{MN}\partial_{M}\M^{KL}\partial_{L}\M_{NK} \\
		&\quad -\frac{1}{7200}\,f^{NQ}{}_{P}f^{MS}{}_{R}\,\M^{PK}\partial_{M}\M_{QK}\,\M^{RL}\partial_{N}\M_{SL}\\
		&\quad +\frac{1}{2}\,g^{-1}\partial_{M}g\,\partial_{N}\M^{MN} + \frac{1}{4}\,\M^{MN}\,g^{-2}\partial_{M}g\,\partial_{N}g + \frac{1}{4}\,\M^{MN}\,\partial_{M}g_{\mu\nu}\,\partial_{N}g^{\mu\nu}\,.
	\end{aligned}
\end{equation}
Finally, $\mathscr{L}_{\rm CS}$ is a Chern-Simons term required to impose the on-shell duality between scalar and vector fields, given by
\begin{equation}
	\begin{aligned}
		\mathscr{L}_{\rm CS} &= \frac{1}{2}\,\varepsilon^{\mu\nu\rho}\,\bigg(F_{\mu\nu}{}^{M}B_{\rho\,M}-f_{KL}{}^{N}\,\partial_{\mu}A_{\nu}{}^{K}\partial_{N}A_{\rho}{}^{L}-\frac{2}{3}\,f^{N}{}_{KL}\,\partial_{M}\partial_{N}A_{\mu}{}^{K}A_{\nu}{}^{M}A_{\rho}{}^{L}\\
		&\quad -\frac{1}{3}\,f_{MKL}\,f^{KP}{}_{Q}\,f^{LR}{}_{S}\,A_{\mu}{}^{M}\partial_{P}A_{\nu}{}^{Q}\partial_{R}A_{\rho}{}^{S}\bigg),
	\end{aligned}
\end{equation}
where $F_{\mu\nu}{}^{M}$ is the covariant field strength of $A_{\mu}{}^{M}$ (see eq. (2.26) of ref.~\cite{Hohm:2014fxa}).

\section{Consistent truncations to 3-dimensional \texorpdfstring{${\cal N}=16$}{N=16} gauged supergravity} \label{s:TruncationReview}
The \EE exceptional field theory is well suited to the construction of consistent truncation of type~II and 11d supergravities to three-dimensional maximal supergravity. These truncations arise as generalised Scherk-Schwarz reductions, described by an \EE-valued twist matrix $U_{M}{}^{\fl{M}}$ and a scale factor $\rho$. The truncation Ansätze are as follows~\cite{Galli:2022idq}:
\begin{equation} \label{eq:GSSansatz}
	\begin{aligned}
		g_{\mu\nu}(x,Y)&=\rho(Y)^{-2}\,\mathring{g}_{\mu\nu}(x)\,,\\
		\M_{MN}(x,Y)&=U_{M}{}^{\fl{M}}(Y)\,U_{N}{}^{\fl{N}}(Y)\,\M_{\fl{MN}}(x)\,,\\
		A_{\mu}{}^{M}(x,Y)=&\rho(Y)^{-1}\,\UI_{\fl{M}}{}^{M}(Y)\,\A_{\mu}{}^{\fl{M}}(x)\,,\\
		B_{\mu\,M}(x,Y)&=\Sigma_{\fl{M}\,M}(Y)\,\A_{\mu}{}^{\fl{M}}(x)\,,
	\end{aligned}
\end{equation}
with
\begin{equation}
	\Sigma_{\fl{M}\,M}=\frac{\rho^{-1}}{60}\,f_{\fl{M}}{}^{\fl{PQ}}\,\UI_{\fl{P}P}\,\partial_{M}\UI_{\fl{Q}}{}^{P}\,.
\end{equation}
The fields with flat indices $\fl{M}$ (in the $\mathbf{248}$ representation of \EE) belong to the three-dimensional theory; all the dependence on the internal manifold is factored out in $U_{M}{}^{\fl{M}}$ and $\rho$. Note that with the definition of $\Sigma_{\fl{M}\,M}$ the condition~\eqref{eq:sectionconstraint2} is automatically satisfied.

The Ansätze~\eqref{eq:GSSansatz} describe a \textit{consistent} truncation if the following condition is satisfied:
\begin{equation} \label{eq:GenLeibniz}
	\gL_{\ff{U}_{\fl{M}}} {\cU}_{\fl{N}}{}^M = X_{\fl{M}\fl{N}}{}^{\fl{P}}\, \cU_{\fl{P}}{}^M\,,
\end{equation}
with $\cU_{\fl{M}}{}^M = \rho^{-1} \UI_{\fl{M}}{}^M$, $\ff{U}_{\fl{M}} = (\cU_{\fl{M}},\,\Sigma_{\fl{M}})$ and $X_{\fl{M}\fl{N}}{}^{\fl{P}}$ a constant tensor. This condition ensures that the factorised form of the Ansätze~\eqref{eq:GSSansatz} is preserved by the generalized Lie derivative, \textit{e.g.}
\begin{equation} \label{eq:GSSexample}
	\begin{split}
		D_{\mu}\M(x,Y) &= U_{M}{}^{\fl{M}}(Y)\,U_{N}{}^{\fl{N}}(Y)\,{\cal D}_{\mu}\M_{\fl{MN}}(x)\,,\\
		{\cal L}_{(A_{\mu},B_{\nu})}A_{\mu}(x,Y) &= \rho(Y)^{-1}\,\UI_{\fl{M}}{}^{M}(Y)\,\llbracket\A_{\mu},\A_{\nu}\rrbracket^{\fl{M}}(x)\,,
	\end{split}
\end{equation}
where 
\begin{equation} \label{eq:3dCovDeriv}
	{\cal D}_{\mu}\M_{\fl{MN}} = \partial_{\mu}\M_{\fl{MN}} + 2\,\A_{\mu}{}^{\fl{P}}X_{\fl{P}(\fl{M}}{}^{\fl{Q}}\M_{\fl{N})\fl{Q}}\,, \quad {\rm and} \quad \llbracket\A_{\mu},\A_{\nu}\rrbracket^{\fl{M}}=X_{\fl{PQ}}{}^{\fl{M}}\A_{\mu}{}^{\fl{P}}\A_{\nu}^{\fl{Q}} \,.
\end{equation}

Thus, for a given twist matrix $U_{M}{}^{\fl{M}}$ satisfying the consistency condition~\eqref{eq:GenLeibniz}, the action~\eqref{eq:ExFTaction} reduces to three-dimensional ${\cal N}=16$ gauged supergravity \cite{Nicolai:2000sc,Nicolai:2001sv}. The constant tensor $X_{\fl{M}\fl{N}}{}^{\fl{P}}$ plays the role of the embedding tensor of the three-dimensional gauged supergravity, as is clear from its appearance in \eqref{eq:3dCovDeriv} in the gauge-covariant derivative and the gauge algebra of the vector fields. From eq. \eqref{eq:GenLeibniz}, it has the following expression in terms of the twist matrix and $\rho$:
\begin{equation}
	\begin{split}
		X_{\fl{MN}}{}^{\fl{P}}&=-\rho^{-1}\,\Gamma_{\fl{MN}}{}^{\fl{P}} + \rho^{-1}\,f^{\fl{P}}{}_{\fl{NQ}}\,f^{\fl{QK}}{}_{\fl{L}}\,\Gamma_{\fl{KM}}{}^{\fl{L}}-\frac{1}{60}\,\rho^{-1}\,f^{\fl{PK}}{}_{\fl{N}}\,f_{\fl{ML}}{}^{\fl{Q}}\,\Gamma_{\fl{KQ}}{}^{\fl{L}}\\
		&\quad-\frac{1}{2}\,\rho^{-1}\,f^{\fl{P}}{}_{\fl{NQ}}\,f^{\fl{QK}}{}_{\fl{M}}\,\Gamma_{\fl{RK}}{}^{\fl{R}}+\left(\delta_{\fl{M}}{}^{\fl{K}}\delta_{\fl{N}}{}^{\fl{P}}-\frac{1}{2}\,f_{\fl{M}}{}^{\fl{LK}}f_{\fl{NL}}{}^{\fl{P}}\right)\,\xi_{\fl{K}}\,.
	\end{split}
\end{equation}
Here, we defined the $\EE$ current $\Gamma_{\fl{MN}}{}^{\fl{P}}=\UI_{\fl{M}}{}^{K}\UI_{\fl{N}}{}^{L}\partial_{K}U_{L}{}^{\fl{P}}$ and the trombone gauging
\begin{equation} \label{eq:Trombone}
	\begin{split}
		\xi_{\fl{M}} 
		&= 2\,U_{\fl{M}}{}^{N}\partial_{N}\rho^{-1} - \rho^{-1}\, \Gamma_{\ov{N}\ov{M}}{}^{\ov{N}} \,.
	\end{split}
\end{equation}
The embedding tensor is most nicely expressed once projected on the adjoint representation:
\begin{equation} \label{eq:symembeddingtensor}
	\begin{split}
		X_{\fl{MN}} = -\frac{1}{60}\,X_{\fl{MP}}{}^{\fl{Q}}\,f_{\fl{NQ}}{}^{\fl{P}} &= - 2\,\rho^{-1}\,\Gamma_{(\fl{MN})} - \rho^{-1}\,\Gamma_{\fl{P}(\fl{M}}{}^{\fl{Q}}\,f_{\fl{N})\fl{Q}}{}^{\fl{P}} + \frac12 f_{\ov{M}\ov{N}}{}^{\ov{P}}\, \xi_{\ov{P}} \\
		&= - 14\,\rho^{-1}\,(\mathbb{P}_{3875})_{\fl{MN}}{}^{\fl{PQ}}\,\Gamma_{\fl{PQ}} - \frac{1}{4}\,\rho^{-1}\,\eta_{\fl{MN}}\,\Gamma_{\fl{P}}{}^{\fl{P}} - \frac12 f_{\ov{M}\ov{N}}{}^{\ov{P}}\, \xi_{\ov{P}} \,,
	\end{split}
\end{equation}
with the projection of the current $\Gamma_{\fl{MN}} = - \tfrac{1}{60}\,\Gamma_{\fl{MP}}{}^{\fl{Q}}\,f_{\fl{NQ}}{}^{\fl{P}}$. In the following, we will focus on gaugings with $\xi_{\fl{M}}=0$.

The three-dimensional action follows from inserting the Ansätze~\eqref{eq:GSSansatz} in the ExFT action~\eqref{eq:ExFTaction}. The expressions of the kinetic and Chern-Simons terms result immediately from eq.~\eqref{eq:GSSexample}:
\begin{equation}
	\begin{split}
		\mathscr{L}_{\rm kin} &= \frac{1}{240}\,D_{\mu}\M_{MN}D^{\mu}\M^{MN} \underset{\rm gSS}{=} \frac{\rho^{2}}{240}\,{\cal D}_{\mu}\M_{\fl{MN}}{\cal D}^{\mu}\M^{\fl{MN}}, \\
		\mathscr{L}_{\rm CS} &\underset{\rm gSS}{=} -\rho^{-1}\,\varepsilon^{\mu\nu\rho}\,X_{\fl{MN}}\,\A_{\mu}{}^{\fl{M}} \left(\partial_{\nu}\A_{\rho}{}^{\fl{N}}-\frac{1}{3}\,\llbracket\A_{\nu},\A_{\rho}\rrbracket^{\fl{N}}\right)\,.
	\end{split}
\end{equation}
However, the case of the potential term $\mathscr{L}_{\rm int}$, leading to a potential $V$ for the three-dimensional scalars $\M_{\fl{MN}}$, is more subtle. We expect $V$ to be quadratic in the embedding tensor, which makes the identification of $X_{\fl{MN}}$ more complicated. Moreover, the potential of three-dimensional ${\cal N}=16$ gauged supergravity in terms of the embedding tensor is unknown, with currently the potential only expressed in terms of the fermion shift matrices of the gauged supergravity \cite{Nicolai:2000sc,Nicolai:2001sv}. Thus, in the following, we will follow the truncation procedure carefully and thereby construct the potential of the three-dimensional ${\cal N}=16$ gauged supergravity in terms of the embedding tensor.

\section{Deriving the potential of ${\cal N}=16$ gauged supergravity} \label{s:Potential}
Here we will use the $\EE$ ExFT Lagrangian and the consistent truncation Ansatz \eqref{eq:GSSansatz} to derive the potential of three-dimensional ${\cal N}=16$ gauged supergravity \cite{Nicolai:2000sc,Nicolai:2001sv} in terms of the embedding tensor. Not only is this an interesting application of ExFT, relying on purely bosonic considerations and bypassing the usual construction of the gauged supergravity potential using supersymmetry\footnote{Note that the same strategy was recently used to impressively derive the potential of maximal two-dimensional gauged supergravity \cite{Bossard:2022wvi}, where fermions are extremely poorly understood. As a result, in that case, the bosonic $\En{9}$ ExFT \cite{Bossard:2018utw,Bossard:2021jix} provides the only currently accessible route for computing the gauged supergravity potential.}, but the potential is crucial for finding vacua of the three-dimensional theory and uplifting these in the later parts of this paper.

In order to derive the potential, we adopt the following strategy. We want $\mathscr{L}_{\rm int}$ \eqref{eq:ExFTLint} to reduce to the embedding tensor squared upon inserting the generalised Scherk-Schwarz Ans\"{a}tze \eqref{eq:GSSansatz}. However, as in higher dimensions, \textit{e.g.} \cite{Berman:2012uy,Musaev:2013rq,Blair:2014zba}, this does not have to match identically, but only up to total derivative terms and terms which violate the section condition. The possible boundary terms are
\begin{equation} \label{eq:BoundaryTerms}
	\frac{1}{\sqrt{\dg}} \partial_M \left(\sqrt{\dg} \partial_N \gM^{MN} \right) \quad \text{and} \quad \frac{1}{\sqrt{\dg}} \partial_M \left( \gM^{MN} \partial_N \sqrt{\dg}\right) \,.
\end{equation}
However, since the ${\cal N}=16$ supergravity potential will be given by terms quadratic in $X_{\fl{MN}}{}^{\fl{P}}$, we do not want the boundary terms to involve any double derivative terms. Therefore, the two boundary terms \eqref{eq:BoundaryTerms} can only appear via the combination
\begin{equation} \label{eq:BoundaryTermCombination}
	\frac{1}{\sqrt{\dg}}\,\partial_{M}\Big(\sqrt{\dg}\,\partial_{N}\M^{MN}+\frac{4}{3}\,\M^{MN}\partial_{N}\sqrt{\dg}\Big) \,.
\end{equation}

Now, we first insert the generalized Sherk-Schwarz Ansätze \eqref{eq:GSSansatz} into the expression of $\mathscr{L}_{\rm int}$~\eqref{eq:ExFTLint} with the additional total derivative \eqref{eq:BoundaryTermCombination}
\begin{equation} \label{eq:Lintexpansion}
	\begin{aligned}
		\mathscr{L}_{\rm int} &+ \frac{a}{\sqrt{\vert g\vert}}\,\partial_{M}\Big(\sqrt{\vert g\vert}\,\partial_{N}\M^{MN}+\frac{4}{3}\,\M^{MN}\partial_{N}\sqrt{\vert g\vert}\Big)\\
		&\underset{\rm gSS}{=} -\frac{1}{2}\,\M^{\fl{MN}}\,\Gamma_{\fl{MK}}\Gamma_{\fl{N}}{}^{\fl{K}} -\M^{\fl{MN}}\,\Gamma_{\fl{MK}}\Gamma^{\fl{K}}{}_{\fl{N}} -\frac{1}{2}\,\Gamma_{\fl{MN}}\Gamma^{\fl{NM}} - \frac{1}{2}\,\M^{\fl{MN}}\M^{\fl{KL}}\,\Gamma_{\fl{MK}}\Gamma_{\fl{NL}} \\
		&\ \quad - \frac{1}{2}\,\M^{\fl{MN}}\M^{\fl{KL}}\,\Gamma_{\fl{MK}}\Gamma_{\fl{LN}} +\frac{3}{2}\left(a-1\right)\,\M^{\fl{MN}}\Gamma_{\fl{MN}}{}^{\fl{K}}\Gamma_{\fl{LK}}{}^{\fl{L}}-\frac{a}{2}\,\M^{\fl{MN}}\Gamma_{\fl{KM}}{}^{\fl{K}}\Gamma_{\fl{LN}}{}^{\fl{L}}\\
		&\ \quad + \left(\frac{1}{2}-a\right)\,\M^{\fl{MN}}\Gamma_{\fl{KM}}{}^{\fl{L}}\Gamma_{\fl{LN}}{}^{\fl{K}}+\,\M^{\fl{MN}}\Gamma_{\fl{MK}}{}^{\fl{L}}\Gamma_{\fl{LN}}{}^{\fl{K}}-\frac{1}{2}\,\M^{\fl{MN}}\M^{\fl{KL}}f_{\fl{QL}}{}^{\fl{P}}\Gamma_{\fl{PN}}{}^{\fl{Q}}\Gamma_{\fl{MK}}\,.
	\end{aligned}
\end{equation}

Secondly, knowing that the ${\cal N}=16$ potential must be quadratic in $X_{\fl{MN}}$, we consider the most general quadratic function in $X_{\fl{MN}}$ and develop it in \EE currents using eq.~\eqref{eq:symembeddingtensor}. Note that in higher dimensions, there exist only two quadratic combinations of the embedding tensor:
\begin{equation}
	X_{\fl{M}\fl{N}}{}^{\fl{P}} X_{\fl{Q}\fl{P}}{}^{\fl{N}} \gM^{\fl{MQ}} \qquad \text{and} \qquad X_{\fl{M}\fl{N}}{}^{\fl{P}} X_{\fl{Q}\fl{R}}{}^{\fl{S}} \gM^{\fl{MQ}} \gM^{\fl{N}\fl{R}} \gM_{\fl{P}\fl{S}} \,,
\end{equation}
or equivalently with eq.~\eqref{eq:symembeddingtensor}
\begin{equation}
	X_{\fl{MN}}X_{\fl{PQ}}\,\M^{\fl{MP}}\eta^{\fl{NQ}} \qquad {\rm and} \qquad X_{\fl{MN}}X_{\fl{PQ}}\,\M^{\fl{MP}}\M^{\fl{NQ}}.
\end{equation}
However, in three dimensions, because the vector fields transform in the adjoint representation of $\EE$, we can write two additional terms
\begin{equation} \label{eq:ExtraX2Terms}
	X_{\fl{MN}} X_{\fl{PQ}}\, \eta^{\fl{MP}} \eta^{\fl{NQ}} \qquad \text{and} \qquad X_{\fl{MN}} X_{\fl{PQ}}\, \eta^{\fl{MN}} \eta^{\fl{PQ}} \,,
\end{equation}
where the second term corresponds to the square of the singlet part of the embedding tensor.

It turns out that for gaugings that arise from consistent truncations, these two terms \eqref{eq:ExtraX2Terms} can be related to the square of the trombone tensor. We can square eq.~\eqref{eq:symembeddingtensor} to derive an expression for $X_{\ov{MN}}X^{\ov{MN}}$ in terms of the current $\Gamma$. We find
\begin{equation}\label{eq:XXwitheta}
	X_{\ov{MN}}X^{\ov{MN}} = 21 \rho^{-2}\, \Gamma_{\ov{MN}}\, \Gamma^{\ov{NM}} + 19 \rho^{-2}\left(\Gamma_{\ov M}{}^{\ov M}\right)^2 - 15\, \xi_{\ov{M}}\, \xi^{\ov{M}} \,.
\end{equation}
Similarly, squaring the trombone tensor $\xi$ gives
\begin{equation}\label{eq:xixiwitheta}
	\xi_{\ov M}\,\xi^{\ov M}=\rho^{-2}\left(\Gamma_{\ov M}{}^{\ov M}\Gamma_{\ov N}{}^{\ov N}+\Gamma_{\ov{MN}}\Gamma^{\ov{NM}}\right) \,.
\end{equation}
We can now deduce
\begin{equation}
	X_{\ov{MN}} \,X^{\ov{MN}}= 6\, \xi_{\ov M} \, \xi^{\ov M} - 2\, \rho^{-2}\left(\Gamma_{\ov M}{}^{\ov M}\right)^2 \,,
\end{equation}
which can be expressed entirely in terms of $X_{\fl{M}\fl{N}}$ after tracing eq.~\eqref{eq:symembeddingtensor}:
\begin{equation}\label{eq:reltrX}
	X_{\ov{MN}}\,X^{\ov{MN}} = 6\, \xi_{\ov M}\, \xi^{\ov M} - \frac1{1922}\left(X_{\ov M}{}^{\ov M}\right)^2 \,. 
\end{equation}
Thus, \eqref{eq:reltrX} is a constraint on the embedding tensor of three-dimensional supergravity which must be satisfied in order for it to have an uplift to 10-/11-dimensional supergravity (see appendix~\ref{app:rulingoutgauging} for an example). Note that this constraint is not implied by the quadratic constraint. To see this, simply observe that the theory with $\En{8}$ gauging has $X_{\ov{MN}}=\eta_{\ov{MN}}$ and $\xi_{\fl{M}} = 0$ \cite{Nicolai:2000sc,Nicolai:2001sv}, violating \eqref{eq:reltrX}.

Since we are focusing on gaugings with vanishing trombone $\xi_{\fl{M}} = 0$, \eqref{eq:reltrX} implies that the two terms in \eqref{eq:ExtraX2Terms} are proportional and we only need to consider one of these terms in the gauged supergravity potential. Thus, the most general Ansatz for the supergravity potential, for gaugings that have an uplift to 10-/11-dimensional supergravity\footnote{If we do not require the three-dimensional gauged supergravity to arise from a consistent truncation, \eqref{eq:reltrX} may be violated and we need to include both terms of \eqref{eq:ExtraX2Terms}. However, here we are not interested in such gaugings.}, is
\begin{equation} \label{eq:Vexpansion}
	\begin{aligned}
		-V &= X_{\fl{MN}}X_{\fl{PQ}}\left(\alpha\,\M^{\fl{MP}}\M^{\fl{NQ}}+\beta\,\M^{\fl{MP}}\eta^{\fl{NQ}}+\delta\,\eta^{\fl{MP}}\eta^{\fl{NQ}}
		\right) \\
		&\underset{\rm gSS}{=} \rho^{-2}\bigg(14\,\alpha\,\Gamma_{\fl{MN}}\,\Gamma_{\fl{PQ}}\,\M^{\fl{MP}}\M^{\fl{NQ}}+14\,\alpha\,\Gamma_{\fl{MN}}\,\Gamma_{\fl{PQ}}\,\M^{\fl{MQ}}\M^{\fl{NP}}\\
		&\ \quad +2\,\beta\,\M^{\fl{MN}}\,\Gamma_{\fl{MK}}\Gamma^{\fl{K}}{}_{\fl{N}}+\beta\,\M^{\fl{MN}}\,\Gamma_{\fl{MK}}\,\Gamma_{\fl{N}}{}^{\fl{K}}+\left(-12\,\alpha+2\,\delta
		\right)\,\Gamma_{\fl{MN}}\,\Gamma^{\fl{MN}}\\
		&\ \quad -2\,\beta\,\M^{\fl{MN}}\Gamma_{\fl{MK}}{}^{\fl{L}}\,\Gamma_{\fl{LN}}{}^{\fl{K}}+\beta\,\M^{\fl{MN}}\Gamma_{\fl{KM}}{}^{\fl{L}}\,\Gamma_{\fl{LN}}{}^{\fl{K}}+\left(7\,\alpha+\frac{\beta}{2}\right)\,\M^{\fl{MN}}\Gamma_{\fl{KM}}{}^{\fl{K}}\,\Gamma_{\fl{LN}}{}^{\fl{L}}\\
		&\ \quad +14\,\alpha\,\M^{\fl{MN}}\M^{\fl{KL}}f_{\fl{QL}}{}^{\fl{P}}\Gamma_{\fl{PN}}{}^{\fl{Q}}\,\Gamma_{\fl{MK}}\bigg)\,.
	\end{aligned}
\end{equation}
Finally, by requiring
\begin{equation}
	\sqrt{|g|} \mathscr{L}_{\rm int} \underset{\rm gSS}{=} - \sqrt{|\mathring{g}|} V \,,
\end{equation}
the parameters in \eqref{eq:Lintexpansion} and \eqref{eq:Vexpansion} are fixed to
\begin{equation}
	\alpha = \frac{1}{28}\,,\quad \beta = \frac{1}{2}\,, \quad \delta=\frac{13}{28}\,, \quad a=1 \,. 
\end{equation}

With the generalized Scherk-Schwarz Ansätze~\eqref{eq:GSSansatz}, the action~\eqref{eq:ExFTaction} becomes
\begin{equation} \label{eq:ExFTactiongSS}
	\begin{aligned}
		S_{\rm ExFT} \underset{\rm gSS}{=} \int\d^{248}Y\,\rho^{-1}\int \d^{3}x & \bigg[\sqrt{\vert\mathring{g}\vert}\bigg(\mathring{R}+\frac{1}{240}\,{\cal D}_{\mu}\M_{\fl{MN}}{\cal D}^{\mu}\M^{\fl{MN}}-V\bigg) \\
		&-\varepsilon^{\mu\nu\rho}\,X_{\fl{MN}}\,\A_{\mu}{}^{\fl{M}} \left(\partial_{\nu}\A_{\rho}{}^{\fl{N}}-\frac{1}{3}\,\llbracket\A_{\nu},\A_{\rho}\rrbracket^{\fl{N}}\right)\bigg],
	\end{aligned}
\end{equation}
with the potential
\begin{equation} \label{eq:3dpotential}
	V = X_{\fl{MN}}\,X_{\fl{PQ}}\left(\frac{1}{28}\,\M^{\fl{MP}}\M^{\fl{NQ}}+\frac{1}{2}\,\M^{\fl{MP}}\eta^{\fl{NQ}}+\frac{13}{28}\,\eta^{\fl{MP}}\eta^{\fl{NQ}}\right).
\end{equation}
As in higher dimensions, $X_{\ov{M}\ov{N}}$ obtained from \eqref{eq:GenLeibniz} automatically satisfies the linear constraint of the 3-dimensional maximal gauged supergravity \cite{Nicolai:2000sc,Nicolai:2001sv}, and the section condition implies the quadratic constraints for $X_{\ov{M}\ov{N}}$.

\section{Adding fluxes to consistent truncations} \label{s:AddFlux}
In order to construct the consistent truncation around the ${\cal N}=(4,4)$ AdS$_3 \times S^3 \times S^3 \times S^1$ vacuum of IIB supergravity, a promising starting point is the dyonic $S^3 \times S^3$ truncation of IIB supergravity constructed in \cite{Inverso:2016eet}, and further reducing this on $S^1$. However, this will not give the correct AdS$_3$ vacuum since the truncation is missing the required 7-form flux on $S^3 \times S^3 \times S^1$. Indeed, the 3-dimensional gauged supergravity that would be obtained this way does not have any AdS$_3$ vacua. We can remedy this situation by defining a new consistent truncation by adding a 7-form flux to the one obtained from the $S^3 \times S^3$ reduction constructed in $\En{7}$ ExFT.

This motivates the following question: Given a consistent truncation, \textit{i.e.} $\cU_{\fl{M}}{}^M$, satisfying \eqref{eq:GenLeibniz}, when can we add a new flux component of string theory to the compactification to obtain a new consistent truncation? Adding a new flux component to the truncation is equivalent to twisting the generalised frame as follows
\begin{equation} \label{eq:Twisting}
	\cU_{\fl{M}}{}^M \longrightarrow \cU'_{\fl{M}}{}^M = \cU_{\fl{M}}{}^N \exp(C)_N{}^M \,,
\end{equation}
where $C$ denotes the $\EE$ generator corresponding to the potential we want to add to the compactification. The effect of the twist \eqref{eq:Twisting} is that the generalised Lie derivative of $\cU'$ satisfies
\begin{equation}
	\gL_{\ff{U}'_{\fl{M}}} \cU'_{\fl{N}}{}^M = \left( \gL_{\ff{U}_{\fl{M}}} \cU_{\fl{N}}{}^N + F_{PQ}{}^N\, \cU_{\fl{M}}{}^P\, \cU_{\fl{N}}{}^Q \right) \exp(C)_N{}^M \,,
\end{equation}
where $F_{MN}{}^P$ is a tensor in the $\mathbf{1} \oplus \mathbf{248} \oplus \mathbf{3875}$ of $\EE$, \textit{i.e.} the same representation as the embedding tensor, corresponding to the field strength of the potential $C$ in \eqref{eq:Twisting}. Using \eqref{eq:GenLeibniz}, we now have
\begin{equation}
	\gL_{\ff{U}'_{\fl{M}}} \cU'_{\fl{N}}{}^M = \left( X_{\fl{M}\fl{N}}{}^{\fl{P}} + \rho^{-1}\, F_{\fl{M}\fl{N}}{}^{\fl{P}} \right) \cU'_{\fl{P}}{}^M \,,
\end{equation}
where we defined
\begin{equation} \label{eq:FlatF}
	F_{\fl{M}\fl{N}}{}^{\fl{P}} \equiv \UI_{\fl{M}}{}^M\, \UI_{\fl{N}}{}^N\, U_P{}^{\fl{P}}\, F_{MN}{}^P \,,
\end{equation}
and $X_{\fl{M}\fl{N}}{}^{\fl{P}}$ is already constant. Therefore, we have a consistent truncation if and only if $\rho^{-1}\, F_{\fl{M}\fl{N}}{}^{\fl{P}}$ is constant.

A particularly simple way of having constant $\rho^{-1}\, F_{\fl{M}\fl{N}}{}^{\fl{P}}$ is to switch on fluxes which are stabilised by the twist matrix $U_M{}^{\fl{M}}$, which typically only lives in a subgroup $G \subset \EE$. Therefore, in this case, we can simply tune the $G$-singlet components of the flux $F_{MN}{}^P$ to be proportional to $\rho$ to obtain a new consistent truncation.

\subsection{Adding fluxes to the $S^3$ truncation of 6-dimensional supergravity} \label{s:S3Flux}
As a warm-up for the $S^3 \times S^3 \times S^1$ truncation, let us demonstrate this methodology for the consistent truncation of ${\cal N}=(1,1)$ 6-dimensional supergravity on $S^3$, which was constructed in \cite{Eloy:2021fhc}. As discussed in \cite{Eloy:2021fhc}, the consistent truncation of ${\cal N}=(1,1)$ 6-dimensional supergravity to 3-dimensional half-maximal gauged supergravity can be described using the $\SO{8,4}$ ExFT \cite{Hohm:2017wtr,Samtleben:2019zrh}. On the other hand, the consistent truncation on $S^3$ is conveniently described by twist matrices living in $\SL{4} \simeq \SO{3,3}$ \cite{Lee:2014mla,Hohm:2014qga,Baguet:2015iou} with the embedding $\SO{3,3} \subset \SO{4,4} \subset \SO{8,4}$. However, the twist matrix in \cite{Eloy:2021fhc} differs from this $\SO{3,3}$ twist matrix \cite{Lee:2014mla,Hohm:2014qga,Baguet:2015iou} by an additional parameter, $\lambda$, which gives rise to an external 3-form flux or, equivalently, a new internal 3-form flux.

The construction of the the $\SL{4} \simeq \SO{3,3}$ twist matrix describing the consistent truncation on $S^3$ is detailed in appendix~\ref{s:SLn}. To demonstrate our methodology, we will now show how the parameter $\lambda$ introduced in \cite{Eloy:2021fhc} can be obtained by the twisting procedure described above. Thus, we want to consider the $\SO{3,3} \subset \SO{4,4}$ twist matrix corresponding to $S^3$ \eqref{eq:SLnTwist} and add a 3-form flux via a $\SO{4,4}$ twist. We begin by decomposing $\SO{4,4} \rightarrow \SO{3,3} \times \SO{1,1}$, such that
\begin{equation}
	\begin{split}
		\mathbf{8} &\rightarrow \mathbf{6_0} \oplus \mathbf{1_2} \oplus \mathbf{1_{-2}} \,, \\
		\mathbf{28} &\rightarrow \mathbf{15_0} \oplus \mathbf{6_2} \oplus \mathbf{6_{-2}} \oplus \mathbf{1_0} \,.
	\end{split}
\end{equation}
Correspondingly, we write a $\SO{4,4}$ vector $V^M$ as
\begin{equation}
	V^M = \left( V^A,\, V^z,\, V^{\bar{z}} \right) \,,
\end{equation}
where $A = 1, \ldots, 6$ labels the vector of $\SO{3,3}$ and $z$, $\bar{z}$ label the $\mathbf{1_2}$ and $\mathbf{1_{-2}}$, respectively.

We denote by $U_A{}^{\ov{A}}$ the $\SO{3,3}$ twist matrix corresponding to the $S^3$ truncation constructed in \cite{Lee:2014mla,Baguet:2015iou} and obtained from \eqref{eq:SLnFrame}, \eqref{eq:rho}, \eqref{eq:SLnTwist}. Then, we can add a 3-form potential by twisting with the $\SO{4,4}$ generator
\begin{equation} \label{eq:C2S3example}
	(e^C)^{MN} = \eta^{MN} + C_A \left( t^A \right)^{MN} \,,
\end{equation}
with $C_A$ an element of the $\mathbf{6_{2}}$, and $\left( t^A \right)^{MN}$ the generator corresponding to $\SO{4,4}$ whose only non-zero component is
\begin{equation}
	\left(t^A\right)^{Bz} = \eta^{AB} \,.
\end{equation}
Here $\eta_{MN}$ and $\eta_{AB}$ are the $\SO{4,4}$ and $\SO{3,3}$ invariant metrics, respectively, and are used to raise/lower the corresponding vector indices. Decomposing with respect to the geometric $\SL{3} \times \mathbb{R}^+$ of the $S^3$,  the coordinates on $S^3$ live in the $Y^A = \left( y^i,\, y_i \right)$, $i = 1, 2, 3$, while $C_A = \left( C_i,\, C^i \right)$ naturally contains a 2-form $C^i$. The field strength $H_{(3)} = \partial^A C_A = \partial_i C^i$ is a singlet of $\SO{3,3}$ and thus we see that $\rho^{-1} F_{\ov{M}\ov{N}}{}^{\ov{P}}$ in \eqref{eq:FlatF} is constant and we obtain a consistent truncation with a new 3-form flux. Evaluating the twist matrix
\begin{equation}
	U'_M{}^{\ov{M}} = (e^C)_M{}^N  U_N{}^{\ov{M}} \,,
\end{equation}
explicitly using \eqref{eq:C2S3example} and setting $C^i = \rho^{-1}\,\lambda\, \xi^i$, with $\nabla_i \xi^i = 1$, in the notation of \cite{Eloy:2021fhc}, we obtain precisely the twist matrix used in \cite{Eloy:2021fhc} for the consistent truncation of 6-dimensional ${\cal N}=(1,1)$ supergravity on $S^3$ with $H_3$ flux. 
We will now follow this same procedure in the next section to obtain the consistent truncation of IIB supergravity on $S^3 \times S^3 \times S^1$ with 7-form flux.

\subsection{Adding flux to the $S^3 \times S^3 \times S^1$ truncation of IIB supergravity}
Our strategy for constructing the consistent truncation on $S^3 \times S^3 \times S^1$ with 7-form flux is to embed the $S^3 \times S^3$ truncation of $\En{7}$ ExFT \cite{Inverso:2016eet} into $\EE \rightarrow \En{7} \times \SL{2}$, as described in \cite{Galli:2022idq}. This will give us the consistent truncation on $S^3 \times S^3 \times S^1$ without 7-form flux. Then we add a 7-form flux to this truncation as outlined above.

\subsubsection{The $S^3 \times S^3$ truncation} \label{s:S3S3}
Let us first review the dyonic $S^3 \times S^3$ truncation of IIB supergravity \cite{Inverso:2016eet}, which forms the starting point of our construction. The key step in the construction of \cite{Inverso:2016eet} is that we can use the $\SL{4}$ generalised frame \eqref{eq:SLnFrame} to form a generalised Leibniz parallelisation of $\En{7}$ via the embedding
\begin{equation} \label{eq:SL4Embedding}
	\En{7} \rightarrow \SL{8} \rightarrow \SL{4}_1 \times \SL{4}_2 \times \mathbb{R}^+ \,.
\end{equation}
The fundamental of $\En{7}$ then decomposes as
\begin{equation} \label{eq:56inSL4SL4}
	\mathbf{56} \rightarrow \mathbf{28} \oplus \overline{\mathbf{28}} \rightarrow \left[ \mathbf{\left(6,1\right)_2} \oplus \mathbf{\left(1,6\right)_{-2}} \oplus \mathbf{\left(4,4\right)_0} \right] \oplus \left[ \mathbf{\left(6,1\right)_{-2}} \oplus \mathbf{\left(1,6\right)_{2}} \oplus \mathbf{\left(\overline{4},\overline{4}\right)_0} \right] \,,
\end{equation}
where the first square brackets denote the branching of the $\mathbf{28}$ under $\SL{4}_1 \times \SL{4}_2 \times \mathbb{R}^+$ and the second square brackets that of the $\mathbf{\overline{28}}$. Crucially, for the generalised Leibniz condition \eqref{eq:GenLeibniz} to hold, the coordinates and generalised vector fields corresponding to the $\SL{4}_1$ and $\SL{4}_2$ must be embedded within the $\mathbf{28}$ and $\mathbf{\overline{28}}$, respectively, which we will call ``electric'' and ``magnetic'' coordinates, following \cite{Inverso:2016eet}. Thus, we write the 56 $\En{7}$ coordinates as
\begin{equation}
	Y^M = \left( Y^{\cA\cB} ,\, Y_{\cA\cB} \right) \,,
\end{equation}
with $\cA, \cB = 1, \ldots, 8$ labelling the fundamental of $\SL{8}$, corresponding to the decomposition $\En{7} \rightarrow \SL{8}$. Following the conventions of \cite{Inverso:2016eet}, the coordinates of the two $\SL{4}$ ExFTs are embedded as $Y^{IJ} \subset Y^{\cA\cB}$, with $I, J = 1, 2, 3, 8$, and $Y_{AB} \subset Y_{\cA\cB}$, with $A, B = 4, 5, 6, 7$. Solving the $\SL{4}$ ExFT section condition for $Y^{IJ}$ and $Y_{AB}$ guarantees a solution to the $\En{7}$ ExFT, and we choose the solution where the six physical coordinates of IIB are
\begin{equation} \label{eq:PhysicalCoords}
	y^i = Y^{i8},\, \qquad \tilde{y}_a = Y_{a7} \,, \qquad i = 1, 2, 3 \,, \qquad a = 4, 5, 6 \,.
\end{equation}
This solution of the section condition defines the ``geometric'' $\SL{3}_1 \times \RO \subset \SL{4}_1$ and $\SL{3}_2 \times \RT \subset \SL{4}_2$ subgroups, that will play an important role when adding flux in section~\ref{s:AddFluxS3S3S1}.

We can now use one copy of the $\SL{4}$ frame \eqref{eq:SLnFrame} for each $\SL{4}$ subgroup of \eqref{eq:SL4Embedding} to construct a generalised parallelisation for the full $\En{7}$ ExFT, with an embedding tensor via \eqref{eq:GenLeibniz} given as in~\eqref{eq:SLnEmbeddingTensor} embedded in $\En{7}$ as follows. Under $\En{7} \rightarrow \SL{8}$, the embedding tensor representation branches as
\begin{equation}
	\mathbf{912} \rightarrow \mathbf{36} \oplus \overline{\mathbf{36}} \oplus \mathbf{420} \oplus \overline{\mathbf{420}} \,,
\end{equation}
and we only generate the $\mathbf{36}$ and $\overline{\mathbf{36}}$ components, $\eta_{{\cal AB}}$, $\tilde{\eta}^{{\cal AB}}$. Thus, the $\En{7}$ embedding tensor is given by
\begin{equation}
	\begin{split}
		X_{{\cal AB, CD}}{}^{\cal EF} &= \eta_{\cA[\cC} \delta_{{\cal D}]\cB}^{{\cal EF}} - \eta_{\cB[\cC} \delta_{{\cal D}]\cA}^{{\cal EF}} \,, \\
		X^{{\cal AB}}{}_{{\cal CD}}{}^{\cal EF} &= -\tilde{\eta}^{\cA[{\cal E}}_{\phantom{[\cA]}} \delta_{\cC{\cal D}}^{{\cal F}]{\cal B}} + \tilde{\eta}^{\cB[{\cal E}}_{\phantom{[\cA]}} \delta_{\cC{\cal D}}^{{\cal F}]{\cal A}} \,.
	\end{split}
\end{equation}
Because we are embedding the $\SL{4}_1$ coordinates in the $\mathbf{28}$ but those of $\SL{4}_2$ in the $\overline{\mathbf{28}}$, the twist matrix will generate an embedding tensor in the $\mathbf{36}$ for $\SL{4}_1$ and $\overline{\mathbf{36}}$ for $\SL{4}_2$, whose only non-vanishing components are
\begin{equation}
	\eta_{IJ} = \delta_{IJ} \,, \qquad \tilde{\eta}^{AB} = \delta^{AB} \,.
\end{equation}

We can also rescale their $S^3$'s using their corresponding $\mathbf{R}^+$ generator. Without loss of generality, we can rescale $S^3$ to have radius $\alpha$, using the $\mathbf{R}^+_1$ generator on the frame. The embedding tensor then becomes
\begin{equation}
	\eta_{IJ} = \frac{1}{\alpha} \delta_{IJ} \,, \qquad \tilde{\eta}^{AB} = \delta^{AB} \,.
\end{equation}

\subsubsection{The $S^3 \times S^3 \times S^1$ truncation} \label{s:S3S3S1}
We now construct the consistent truncation on $S^3 \times S^3 \times S^1$ by embedding the $\En{7}$ twist matrix corresponding to the $S^3 \times S^3$ truncation reviewed above in \ref{s:S3S3} in $\EE$ via the branching $\EE \rightarrow \En{7} \times \SL{2}$, such that $\mathbf{248} \rightarrow \mathbf{\left(133,1\right)} \oplus \mathbf{\left(56,2\right)} \oplus \mathbf{\left(1,3\right)}$. As explained in \cite{Galli:2022idq}, we can construct a consistent truncation on $M \times S^1$ to 3-dimensional gauged supergravity by embedding a consistent truncation on $M$ to 4-dimensional supergravity, characterised by an $\En{7}$ twist matrix and $\En{7}$ scalar density $\sigma$, as well as 
the $\SL{2}$ twist matrix given by
\begin{equation}
	v_{\ov{i}}{}^i = \begin{pmatrix}
		\sigma & 0 \\ 0 & \sigma^{-1}
	\end{pmatrix} \,.
\end{equation}
Finally, the $\EE$ scalar density satisfies $\rho = \sigma^2$ to ensure a consistent truncation on $M \times S^1$.

Employing this procedure for $M = S^3 \times S^3$, gives us the consistent truncation on $S^3 \times S^3 \times S^1$. However, this does not have any AdS$_3$ vacua, since we still need to add the 7-form flux. To do this, let us first review the group theory of the $S^3 \times S^3$ truncation, since this will allow us to determine whether the 7-form flux is stabilised by the twist matrix.

\subsubsection{Adding flux} \label{s:AddFluxS3S3S1}

To add a 7-form flux to the above truncation, we need to determine whether the 7-form flux is stabilised by the twist matrix on $S^3 \times S^3 \times S^1$ constructed in section \ref{s:S3S3S1}. Thus, we need to understand whether IIB supergravity admits a 7-form field strength that is a singlet under the two $\SL{4}$ groups in \eqref{eq:SL4Embedding}. To answer this question, we pick a gauge where the 6-form potential lives entirely in $S^3 \times S^3$ but depends on the $S^1$ coordinate, $z$. Therefore, in this gauge choice, the 6-form potential corresponds to an adjoint generator of $\En{7}$, and since the $S^1$ coordinate is an $\En{7}$ singlet, this adjoint generator must be a singlet under $\SL{4}_1 \times \SL{4}_2 \subset \SL{8} \subset \En{7}$ for us to have a consistent truncation.

IIB supergravity contains an S-duality doublet of 6-forms, which we can easily identify using the decomposition $\En{7} \rightarrow \SL{6} \times \SL{2} \times \mathbb{R}^+_{\rm IIB}$, under which the $\En{7}$ adjoint decomposes as
\begin{equation} \label{eq:AdjointSL6}
	\begin{split}
		\mathbf{133} &\rightarrow \mathbf{\left(35,1\right)_0} \oplus \mathbf{\left(1,3\right)_0} \oplus \mathbf{\left(1,1\right)_0} \oplus \mathbf{\left(1,2\right)_{\pm 6}} \\
		& \quad \oplus \mathbf{\left(\overline{15},2\right)_2} \oplus \mathbf{\left(\overline{15},1\right)_{-4}} \oplus \mathbf{\left(15,2\right)_{-2}} \oplus \mathbf{\left(15,1\right)_4} \,.
	\end{split}
\end{equation}
The 6-form doublet corresponds to the $\mathbf{\left(1,2\right)}_{6}$. To understand if one of the 6-forms in the doublet are singlets under $\SL{4}_1 \times \SL{4}_2 \subset \En{7}$, we must decompose $\En{7}$ with respect to the common subgroup of $\SL{4}_1 \times \SL{4}_2$ and $\SL{6} \times \SL{2}$, which is $\SL{3}_1 \times \SL{3}_2 \times \RO \times \RT$, as defined by the embedding of the physical coordinates in \eqref{eq:PhysicalCoords}. Important for us is the correct identification of the $\mathbb{R}^+$ charges in both decompositions. We have, on the one hand,
\begin{equation} \label{eq:SL4SL4}
	\begin{split}
		\En{7} &\rightarrow \SL{8} \rightarrow \SL{4} \times \SL{4} \times \mathbb{R}^+_{\SL{8}} \\
		&\rightarrow \SL{3}_1 \times \SL{3}_2 \times \RO \times \RT \times \mathbb{R}^+_{\SL{8}} \,,
	\end{split}
\end{equation}
and, on the other,
\begin{equation} \label{eq:SL6SL2}
	\begin{split}
		\En{7} &\rightarrow \SL{6} \times \SL{2} \times \mathbb{R}^+_{\rm IIB} \\
		&\rightarrow \SL{3}_1 \times \SL{3}_2 \times \mathbb{R}^+_{\SL{2}} \times \mathbb{R}^+_{\SL{6}} \times \mathbb{R}^+_{\rm IIB} \,.
	\end{split}
\end{equation}
Here, we label the $\mathbb{R}^+$'s with a subscript that refers to the the groups they belong to. The $\mathbb{R}^+$ generators of the two decompositions are related as
\begin{equation} \label{eq:R+charges}
	\begin{split}
		\mathbb{R}^+_{\SL{6}} &= - \frac12 \left( \RO + \RT \right) \,, \\
		\mathbb{R}^+_{\SL{2}} &= \frac14 \left( \RO - \RT + \mathbb{R}^+_{\SL{8}} \right) \,, \\
		\mathbb{R}^+_{\rm IIB} &= \frac12 \left( \RO - \RT - 3\, \mathbb{R}^+_{\SL{8}} \right) \,.
	\end{split}
\end{equation}

The branching of the adjoint \eqref{eq:AdjointSL6} under $\SL{6} \times \SL{2} \times \mathbb{R}^+_{\rm IIB}$ in \eqref{eq:SL6SL2} needs to now be compared with that via $\SL{4}_1 \times \SL{4}_2 \times \mathbb{R}^+_{\SL{8}}$, given by
\begin{equation} \label{eq:133DecompSL4SL4}
	\begin{split}
		\mathbf{133} &\rightarrow \mathbf{63} \oplus \mathbf{70} \\
		&\rightarrow \mathbf{\left(15,1\right)_0} \oplus \mathbf{\left(1,15\right)_0} \oplus \mathbf{\left(1,1\right)_0} \oplus \mathbf{\left(4,\overline{4}\right)_2} \oplus \mathbf{\left(\overline{4},4\right)_{-2}} \\
		& \quad \oplus \mathbf{\left(1,1\right)_{-4}} \oplus \mathbf{\left(1,1\right)_{4}} \oplus \mathbf{\left(4,\overline{4}\right)_{-2}} \oplus \mathbf{\left(\overline{4},4\right)_{2}} \oplus \mathbf{\left(6,6\right)_0} \,.
	\end{split}
\end{equation}
Using \eqref{eq:R+charges}, we can now identify each of the singlet generators $\left(1,1\right)_{\pm4}$ with one of the $\SL{2}$ doublet generators of charge $\mp 6$. Therefore, we find exactly one singlet $\SL{4}_1 \times \SL{4}_2$ generator, which we can write as $t_{1238}$, corresponding to a 6-form potential. The other $\SL{4}_1 \times \SL{4}_2$ singlet generator, $t_{4567}$, only differs by a compact generator and thus does not correspond to a different physical field. Finally, the other elements of the $\SL{2}$ doublet of 6-form potential can be mapped to the $\mathbf{\left(4,\overline{4}\right)_{2}} \oplus \mathbf{\left(\overline{4},4\right)_{-2}}$ generators, specifically $t_7{}^8$ and $t_8{}^7$ and are clearly not $\SL{4}$ singlets.

Therefore, we can add a 7-form flux to the twist matrix using
\begin{equation}
	\cU_{\fl{M}}{}^M \longrightarrow \cU'_{\fl{M}}{}^M = \cU_{\fl{M}}{}^N \exp(C)_N{}^M \,,
\end{equation}
with $C_M{}^N = \lambda\, z\,\rho^{-1}\, t_{1238\,M}{}^N$ and $\lambda$ a numerical parameter corresponding to the amount of 7-form flux. The resulting embedding tensor is most nicely written using the branching
\begin{equation} \label{eq:248decompoinSL8}
	\begin{aligned}
		\EE & \longrightarrow E_{7(7)}\times {\rm SL}(2) \longrightarrow {\rm SL}(8) \times {\rm SL}(2) \\
		\mathbf{248} & \ \tikz{\draw[arrows={->[scale=1.1]}] (0,0) -- (3.65
	,0);}\  (\mathbf{63},\mathbf{1}) \oplus (\mathbf{70},\mathbf{1}) \oplus (\mathbf{28},\mathbf{2}) \oplus (\mathbf{\fl{28}},\mathbf{2}) \oplus (\mathbf{1},\mathbf{3}) \\
		X^{\fl{M}} & \ \tikz{\draw[arrows={->[scale=1.1]}] (0,0) -- (3.65
	,0);}\ \left\{X^{\fl{\cal A}}{}_{\fl{\cal B}},\,X^{\fl{\cal ABCD}},\,X^{\fl{\cal AB}\,\fl{i}},\,X_{\fl{\cal AB}}{}^{\fl{i}},\,X^{\fl{\alpha}}\right\},
	\end{aligned}
\end{equation}
and has the following non-vanishing components under the $\SL{4}_1 \times \SL{4}_2$ subgroups of $\SL{8}$:
\begin{equation} \label{eq:embeddingtensor@origin}
	\begin{split}
		X_{(\mathbf{63},\mathbf{1});(\mathbf{28},\mathbf{2})}:&
		\begin{cases}
			\displaystyle X_{\fl{I}}{}^{\fl{J}}{}_{;\fl{KL}+}=-\frac{1}{\alpha}\,\delta_{\fl{I}[\fl{K}}\delta_{\fl{L}]}{}^{\fl{J}},\\[7pt]
			\displaystyle X_{\fl{A}}{}^{\fl{I}}{}_{;\fl{BJ}+}=-\frac{1}{2\alpha}\,\delta_{\fl{AB}}\delta_{\fl{J}}{}^{\fl{I}}\,,\\
		\end{cases} \\[5pt]
		X_{(\mathbf{63},\mathbf{1});(\mathbf{\fl{28}},\mathbf{2})}:&
		\begin{cases}
			\displaystyle X_{\fl{A}}{}^{\fl{B}}{}_{;}{}^{\fl{CD}}{}_{+}=-\,\delta_{\fl{A}}{}^{[\fl{C}}\delta^{\fl{D}]\fl{B}}\,,\\[7pt]
			\displaystyle X_{\fl{A}}{}^{\fl{I}}{}_{;}{}^{\fl{BJ}}{}_{+}=-\frac{1}{2}\,\delta_{\fl{A}}{}^{\fl{B}}\delta^{\fl{IJ}}\,,\\
		\end{cases} \\[5pt]
		X_{(\mathbf{70},\mathbf{1});(\mathbf{1},\mathbf{3})}:&\  X_{\fl{IJKL};++} = -\frac{\lambda}{12\sqrt{6}}\,\varepsilon_{\fl{IJKL}}\,.
	\end{split}
\end{equation}
With this embedding tensor, the potential~\eqref{eq:3dpotential} has a vacuum at the scalar origin when $\lambda=\pm4\sqrt{3}\,\sqrt{1+(1/\alpha^{2})}$. The sign changes the chirality of the Killing spinors and, therefore, without loss of generality, we will choose $\lambda = 4\sqrt{3}\,\sqrt{1+(1/\alpha^{2})}$ from here onwards.

\subsection{Adding fluxes to other consistent truncations}
We can also try to apply the same procedure we described here to turn on fluxes in other consistent truncations. One set of natural candidates are the $S^1$ reductions of the consistent truncations to dyonically gauged 4-dimensional maximal supergravities \cite{Guarino:2015jca,Inverso:2016eet}. These 4-dimensional truncations arise from compactifications of 10-dimensional supergravity on $S^6$, $S^5\times S^1$, $S^4\times S^2$, with the 10-d theory corresponding to (massive) IIA/IIB when the spheres have even/odd dimensions, respectively. The corresponding twist matrices for $S^6$, $S^5\times S^1$, $S^4\times S^2$ belong to $\SL{8-p} \times \SL{p}$, with $p = 1,\, 2,\, 3$ for the respective cases, and are constructed as in appendix \ref{s:SLn}.

Let us see whether we can turn on a similar flux as in the case of $S^3 \times S^3 \times S^1$ for these cases. Here we will consider the case of a flux which has one leg on $S^1$, so as to require the truncation to 3 dimensions. In this case, we can always choose a gauge where the potential is fully on $S^{7-p} \times S^{p-1}$ and depends on the $S^1$ coordinate. Turning on such a flux only leads to a consistent truncation if we can find singlets in the $\mbf{133}$ of $\En{7}$ under $\SL{8-p} \times \SL{p}$ for the respective cases of $S^{7-p} \times S^{p-1}$, $p = 1,\, 2,\, 3$, that correspond to $p$-form potential. And, indeed for all these cases, the $\mbf{133}$ only contains exactly one singlet, corresponding to the $\mathbb{R}^+$ generator that commutes with $\SL{8-p} \times \SL{p}$.

However, this $\mathbb{R}^+$ generator cannot correspond to a form potential. If it did, there would also have been a corresponding compact generator in the same representation (with possibly different $\mathbb{R}^+$ chargse), just as in \eqref{eq:133DecompSL4SL4}, associated to the ``dual'' potential. Since we only have one $\SL{8-p} \times \SL{p}$ singlet, this cannot be a $p$-form potential and, therefore, turning on new fluxes with legs on $S^1$ will not lead to consistent truncation for $S^{7-p} \times S^{p-1} \times S^1$, $p = 1,\, 2,\, 3$. This analysis also reveals why the $S^3\times S^3 \times S^1$ case is special: the $\mbf{133}$ contains a 4-form of $\SL{8}$, \textit{i.e.} the $\mathbf{70}$, which naturally contains the relevant singlets under $\SL{4}\times \SL{4}$.

Our analysis does not rule out the possibility of adding fluxes whose only legs are on $S^{7-p} \times S^p$, with none on $S^1$. In this case, we could use the analysis above but look for singlets in the $\mathbf{56} \otimes \mathbf{133}$, corresponding to the fluxes on $S^{7-p} \times S^p$.

\section{Moduli of \texorpdfstring{AdS$_3 \times S^3 \times S^3 \times S^1$}{AdS3xS3xS3xS1}} \label{s:Moduli}

\subsection{The undeformed ${\cal N}=(4,4)$ AdS$_3 \times S^3 \times S^3 \times S^1$ solution}
\paragraph{} At $\lambda=4\sqrt{3}\,\sqrt{1+(1/\alpha^{2})}$, the three-dimensional solution described by the embedding tensor~\eqref{eq:embeddingtensor@origin} corresponds to the consistent truncation of the pure NSNS ten-dimensional solution~\cite{deBoer:1999gea}
\begin{equation} \label{eq:AdS3}
	\begin{split}
		\d s^{2} &= \ell_{\rm AdS}^{2}\,\d s^{2}\left({\rm AdS}_{3}\right) + \alpha^{2}\,\d s^{2}\big(S^{3}\big)+\d s^{2}\big(\ts{S}^{3}\big)+\d z^{2},\\
		H_{(3)} &= 2\left(\ell_{\rm AdS}^{2}\,{\rm Vol}({\rm AdS_{3}})+\alpha^{2}\,{\rm Vol}\big(S^{3}\big)+{\rm Vol}\big(\ts{S}^{3}\big)\right)\,,
	\end{split}
\end{equation}
with $H_{(3)} = \d B_{(2)}$ and
\begin{equation}
	\ell_{\rm AdS}^{2} = \frac{\alpha^{2}}{1+\alpha^{2}}\,.
\end{equation}
The $S^{1}$ coordinate $z$ has periodicity $z\rightarrow z+1$.

\paragraph{}
The AdS$_3$ solution \eqref{eq:AdS3} preserves ${\cal N}=(4,4)$ supersymmetry, with super-isometry group ${\cal G}=D^{1}(2,1;\alpha)_{\rm L}\times D^{1}(2,1;\alpha)_{\rm R}$, \textit{i.e.} two copies of the large ${\cal N}=4$ supergroup. The even part of $D^{1}(2,1;\alpha)$ is isomorphic to ${\rm SL}(2,\mathbb{R})\times{\rm SO}(3)\times\ts{{\rm SO}(3)}$ and the bosonic isometries of the ${\rm AdS}_{3}\times S^{3}\times\ts{S}^{3}$ background are built from the even part of ${\cal G}$. The ${\rm AdS}_{3}$ isometry group is ${\rm SO}(2,2)\simeq{\rm SL}(2,\mathbb{R})_{\rm L}\times{\rm SL}(2,\mathbb{R})_{\rm R}$, and the factors ${\rm SO}(3)_{\rm L}\times{\rm SO}(3)_{\rm R}\times\ts{{\rm SO}(3)}_{\rm L}\times\ts{{\rm SO}(3)}_{\rm R}\simeq{\rm SO}(4)\times\ts{{\rm SO}(4)}$ combine into the isometry groups of the two spheres. Additionally, there is a $\U{1}_F$ symmetry due to the $S^1$ factor in \eqref{eq:AdS3}. The spectrum organizes into multiplets of ${\cal G}$, made out of products of long representations $[\ell,\ts{\ell}]$ of $D^{1}(2,1;\alpha)$ (see appendix~\ref{app:d21a} for a review of the representations of $D^{1}(2,1;\alpha)$), and carrying a charge $n$ under the $\U{1}_F$. Explicitly, the spectrum is~\cite{Eberhardt:2017fsi}
\begin{equation} \label{eq:spectrumorigin}
	\bigoplus_{\ell,\ts{\ell}\geq0,\,n} \Big([\ell,\ts{\ell}]\otimes[\ell,\ts{\ell}]\Big)_n \ \ominus\ [0,0]_{\rm s} \otimes [0,0]_{\rm s} \,.
\end{equation}
In each factor the lowest conformal dimension is
\begin{equation}
	h_{\rm L} = h_{\rm R} = -\frac{1}{2} + \frac{1}{2}\sqrt{1+\frac{4\,\ell(\ell+1)+4\,\alpha^{2}\,\ts{\ell}(\ts{\ell}+1)+\alpha^2(2\pi n)^{2}}{1+\alpha^{2}}}\,,
\end{equation}
with $n$ corresponding to the $S^{1}$ harmonics, and the lowest conformal dimension of $[\ell,\ts{\ell}]\otimes[\ell,\ts{\ell}]$ is then
\begin{equation} \label{eq:confdims3s3s1}
	\Delta_{\ell,\,\ts{\ell},\,n} = h_{\rm L} + h_{\rm R} = - 1 + \sqrt{1+\frac{4\,\ell(\ell+1)+4\,\alpha^{2}\,\ts{\ell}(\ts{\ell}+1)+\alpha^2(2\pi n)^{2}}{1+\alpha^{2}}}\,.
\end{equation}
For $n=0$, the representations get shortened whenever $\ell = \ts{\ell}$. The spectrum at the consistent truncation is then the one of ref.~\cite{Hohm:2005ui}. It corresponds to $\ell=\ts{\ell}=n=0$ in eq.~\eqref{eq:spectrumorigin}:
\begin{equation} \label{eq:CTmultorigin}
	[0,0]_{\rm s} \otimes [\nicefrac{1}{2},\nicefrac{1}{2}]_{\rm s} \ \oplus \ [\nicefrac{1}{2},\nicefrac{1}{2}]_{\rm s} \otimes [0,0]_{\rm s} \ \oplus \ [\nicefrac{1}{2},\nicefrac{1}{2}]_{\rm s} \otimes [\nicefrac{1}{2},\nicefrac{1}{2}]_{\rm s}\,.
\end{equation}
The first two terms build the supergravity multiplet, each of them carrying one spin-2 field, four gravitini, seven vectors and four spin-$\nicefrac{1}{2}$ fields, all massless. The only propagating degrees of freedom are in $[\nicefrac{1}{2},\nicefrac{1}{2}]_{\rm s}\otimes[\nicefrac{1}{2},\nicefrac{1}{2}]_{\rm s}$ (see the details in table 4 of \cite{Hohm:2005ui}).

\subsection{Moduli in gauged supergravity}
Surprisingly, within the 3-dimensional gauged supergravity, the AdS$_3$ vacuum contains a large moduli space, which even has supersymmetric submanifolds. In particular, a simple search already reveals a continuous 11-parameter family of AdS$_3$ solutions, connected to the origin and given by the \EE coset representative
\begin{equation} \label{eq:FullDeformation}
	\begin{split}
		{\cal V}_{\rm def} &= \exp\left[\sqrt{2}\left(\chi_{1}\big(t_{(\mathbf{28},\mathbf{2})}\big)_{23,-}+\chi_{2}\big(t_{(\mathbf{28},\mathbf{2})}\big)_{81,-}+\ts{\chi}_{1}\big(t_{(\mathbf{\fl{28}},\mathbf{2})}\big)_{56,-}+\ts{\chi}_{2}\big(t_{(\mathbf{\fl{28}},\mathbf{2})}\big)_{74,-}\right)\right] \\
		& \times \exp\left[\sqrt{2}\left(\zeta_{1}\big(t_{(\mathbf{\fl{28}},\mathbf{2})}\big)_{23,+}+\zeta_{2}\big(t_{(\mathbf{\fl{28}},\mathbf{2})}\big)_{81,+}+\ts{\zeta}_{1}\big(t_{(\mathbf{28},\mathbf{2})}\big)_{56,+}+\ts{\zeta}_{2}\big(t_{(\mathbf{28},\mathbf{2})}\big)_{74,+}\right)\right] \\
		& \times \exp\bigg[\frac{\Omega}{120}\Big(\big(t_{(\mathbf{63},\mathbf{1})}\big)_{1}{}^{1}-\big(t_{(\mathbf{63},\mathbf{1})}\big)_{2}{}^{2}-\big(t_{(\mathbf{63},\mathbf{1})}\big)_{3}{}^{3}+\big(t_{(\mathbf{63},\mathbf{1})}\big)_{8}{}^{8}\Big)\\
		&\ \qquad+\frac{\ts{\Omega}}{120}\Big(\big(t_{(\mathbf{63},\mathbf{1})}\big)_{4}{}^{4}-\big(t_{(\mathbf{63},\mathbf{1})}\big)_{5}{}^{5}-\big(t_{(\mathbf{63},\mathbf{1})}\big)_{6}{}^{6}+\big(t_{(\mathbf{63},\mathbf{1})}\big)_{7}{}^{7}\Big)\bigg] \\
		& \times \exp\left[\psi\left(\frac{1}{120}\,\big(t_{(\mathbf{63},\mathbf{1})}\big)_{I}{}^{I}-\frac{1}{120}\,\big(t_{(\mathbf{63},\mathbf{1})}\big)_{A}{}^{A}-\frac{1}{2}\,\big(t_{(\mathbf{1},\mathbf{3})}\big)_{1}\right)\right]\,,
	\end{split}
\end{equation}
where we denoted the generators according to the decomposition~\eqref{eq:248decompoinSL8}. The $(\chi_{i},\ts{\chi}_{i},\zeta_{i},\ts{\zeta}_{i})$ deformations excite the $(\mathbf{6},\mathbf{1})_{\mathbf{2}}$ and $(\mathbf{1},\mathbf{6})_{\mathbf{2}}$ generators in the branching~\eqref{eq:56inSL4SL4}. Note that changing the sign in some terms in \eqref{eq:FullDeformation}, \textit{e.g.} changing the factor with $\Omega$ to
\begin{equation}
	\exp\bigg[\frac{\Omega'}{120}\Big(\big(t_{(\mathbf{63},\mathbf{1})}\big)_{1}{}^{1}+\big(t_{(\mathbf{63},\mathbf{1})}\big)_{2}{}^{2}-\big(t_{(\mathbf{63},\mathbf{1})}\big)_{3}{}^{3}-\big(t_{(\mathbf{63},\mathbf{1})}\big)_{8}{}^{8}\Big) \bigg] \,,
\end{equation}
creates yet another flat direction of the potential. This highlights how many flat directions the AdS$_3$ vacuum appears to have.

\subsection{Moduli in 10 dimensions}
The consistent truncation Ansatz we constructed in this paper can be used to uplift the full 11-parameter family of AdS$_3$ vacua of the 3-dimensional gauged supergravity in \eqref{eq:FullDeformation} to IIB supergravity. However, this full 11-parameter solution can quickly become unwieldy. Therefore, here we will instead focus on a few interesting subsets of deformations.

\paragraph{\boldmath $(\chi, \ts{\chi}, \Omega, \ts{\Omega}, \psi)$}
Using Hopf coordinates for both $S^3$'s we can write the 7-parameter deformation involving $(\chi_i, \ts{\chi}_i, \Omega, \ts{\Omega}, \psi)$, $i = 1,2$, as follows:
\begin{equation} \label{eq:chiomegapsi}
	\begin{split}
		\d s^{2} &= \ell_{\rm AdS}^{2} \,\d s^{2}({\rm AdS}_{3}) + \alpha^{2}\,\d\theta^{2} +\frac{\alpha^{2}}{\cos^{2}(\theta)+e^{\Omega}\,\sin^{2}(\theta)}\left( e^{\Omega}\cos^{2}(\theta)\,\d\phi'_{1}{}^{2} + \sin^{2}(\theta)\,\d\phi'_{2}{}^{2} \right) \\
		&\quad +\d\ts{\theta}^{2} + \frac{1}{e^{\ts{\Omega}}\cos^{2}(\ts{\theta})+\sin^{2}(\ts{\theta})} \left(\cos^{2}(\ts{\theta})\,\d\ts{\phi}'_{1}{}^{2} + e^{\ts{\Omega}}\sin^{2}(\ts{\theta})\,\d\ts{\phi}'_{2}{}^{2}\right) + e^{2\psi}\,\d z^{2}\,, \\[10pt]
		H_{(3)} &= 2\,\ell_{\rm AdS}^{2}\,{\rm Vol}({\rm AdS_{3}}) \\
		&\quad + \frac{2\,\alpha^{2}\,e^{\Omega}\cos(\theta)\sin(\theta)}{\left(\cos^{2}(\theta)+e^{\Omega}\,\sin^{2}(\theta)\right)^{2}}\,\d\theta\wedge\d\phi'_{1}\wedge\d\phi'_{2}+ \frac{2\,e^{\ts{\Omega}}\cos(\ts{\theta})\sin(\ts{\theta})}{\left(e^{\ts{\Omega}}\cos^{2}(\ts{\theta})+\sin^{2}(\ts{\theta})\right)^{2}}\,\d\ts{\theta}\wedge\d\ts{\phi}'_{1}\wedge\d\ts{\phi}'_{2}\,,
	\end{split}
\end{equation}
where
\begin{equation}
	\begin{cases}
		\displaystyle \d\phi'_{1} = \d\phi_{1} + \frac{1}{\alpha}\,\chi_{1}\,\d z\,,\\[6pt]
		\displaystyle \d\phi'_{2} = \d\phi_{2} + \frac{1}{\alpha}\,\chi_{2}\,\d z\,,
	\end{cases}\qquad
	\begin{cases}
		\d\ts{\phi}'_{1} = \d\ts{\phi}_{1} + \ts{\chi}_{1}\,\d z\,,\\[6pt]
		\d\ts{\phi}'_{2} = \d\ts{\phi}_{2} + \ts{\chi}_{2}\,\d z\,.
	\end{cases}
\end{equation}
This 7-parameter deformation generically breaks $\SO{3}_{\rm L}\times\SO{3}_{\rm R}\times\ts{\SO{3}}_{\rm L}\times\ts{\SO{3}}_{\rm R}$ to ${\rm U}(1)_{\rm L}\times{\rm U}(1)_{\rm R}\times\ts{{\rm U}(1)}_{\rm L}\times\ts{{\rm U}(1)}_{\rm R}$ and all supersymmetries. The moduli space of the solution is described by the following metric:
\begin{equation} \label{eq:Zam7Param}
	\begin{aligned}
		\d s^{2}_{\rm mod.} &= -\frac{1}{2}\bigg[e^{-2\psi}\left(e^{\Omega}\,\d\chi_{1}^{2}+e^{-\Omega}\,\d\chi_{2}^{2}+e^{-\ts{\Omega}}\,\d\ts{\chi}_{1}^{2}+e^{\ts{\Omega}}\,\d\ts{\chi}_{2}^{2}\right) +2\,\d\psi^{2} + \frac{1}{2}\,\d\Omega^{2}+ \frac{1}{2}\,\d\ts{\Omega}^{2}\bigg] \,.
	\end{aligned}
\end{equation}

\paragraph{}
A particularly interesting family of solutions is obtained in the case $\Omega=\ts{\Omega}=0$, where eq.~\eqref{eq:chiomegapsi} reduces to
\begin{equation} \label{eq:chipsi}
	\begin{split}
		\d s^{2} &= \ell_{\rm AdS}^{2} \,\d s^{2}({\rm AdS}_{3}) +\alpha^{2}\left(\d\theta^{2} + \cos^{2}(\theta)\,\d\phi'_{1}{}^{2} + \sin^{2}(\theta)\,\d\phi'_{2}{}^{2} \right) \\[3pt]
		&\quad +\d\ts{\theta}^{2} + \cos^{2}(\ts{\theta})\,\d\ts{\phi}'_{1}{}^{2} + \sin^{2}(\ts{\theta})\,\d\ts{\phi}'_{2}{}^{2} + e^{2\psi}\,\d z^{2}\,, \\[5pt]
		H_{(3)} &= 2\,\ell_{\rm AdS}^{2}\,{\rm Vol}({\rm AdS_{3}}) + 2\,\alpha^{2}\cos(\theta)\sin(\theta)\,\d\theta\wedge\d\phi'_{1}\wedge\d\phi'_{2}+ 2\cos(\ts{\theta})\sin(\ts{\theta})\,\d\ts{\theta}\wedge\d\ts{\phi}'_{1}\wedge\d\ts{\phi}'_{2}\,.
	\end{split}
\end{equation}
Again, the deformation generically breaks $\SO{3}_{\rm L}\times\SO{3}_{\rm R}\times\ts{\SO{3}}_{\rm L}\times\ts{\SO{3}}_{\rm R}$ to ${\rm U}(1)_{\rm L}\times{\rm U}(1)_{\rm R}\times\ts{{\rm U}(1)}_{\rm L}\times\ts{{\rm U}(1)}_{\rm R}$ and all supersymmetries. The $\chi_i$, $\ts{\chi}_i$, $i = 1, 2$ deformations are analogous to the two-parameter ``flat deformations'' \cite{Guarino:2021hrc} of AdS$_4 \times S^5 \times S^1$ \cite{Guarino:2020gfe,Giambrone:2021zvp,Giambrone:2021wsm}. From the 4-dimensional perspective obtained after reducing on $S^3 \times S^3$, these deformations correspond to turning on Wilson lines for the $\SO{3}_{\rm L}\times\SO{3}_{\rm R}\times\ts{\SO{3}}_{\rm L}\times\ts{\SO{3}}_{\rm R}$ gauge fields along $S^1$. The $\psi$ deformation is even simpler: it just rescales the $S^1$ radius. These $\chi_i$, $\ts{\chi}_i$, $\psi$ deformations are completely analogous to $\mathbb{C}$-structure deformations of $T^2$, with the $\chi_i$, $\ts{\chi}_i$ corresponding to the real part and the $\psi$ deformation to the imaginary part of the $\mathbb{C}$-structure deformation. Note that the $\psi$ deformation does not exist in the otherwise analogous AdS$_4 \times S^5 \times S^1$ vacua in \cite{Guarino:2020gfe,Guarino:2021hrc,Giambrone:2021zvp,Giambrone:2021wsm}, due to the ``S-fold'' identification in that case. This leads to an important difference in the global structure of the moduli space that we will discuss in section \ref{s:Compact}.

These deformations affect the Kaluza-Klein spectrum as follows: the conformal dimension of each physical field (\textit{c.f.} eq.~\eqref{eq:confdims3s3s1}) get shifted by replacing
\begin{equation} \label{eq:confdimshiftchi}
	2\pi n\longrightarrow e^{-\psi} \left[ 2\pi n + \frac{1}{2}\Big((q_{\rm L}+q_{\rm R})\,\frac{\chi_{1}}{\alpha}+(q_{\rm L}-q_{\rm R})\,\frac{\chi_{2}}{\alpha}+ (\ts{q}_{\rm L}+\ts{q}_{\rm R})\,\ts{\chi}_{1} + (\ts{q}_{\rm R}-\ts{q}_{\rm L})\,\ts{\chi}_{2}\Big) \right]\,,
\end{equation}
where the $q$ denote the charges under the different ${\rm U}(1)$.

Eq.~\eqref{eq:confdimshiftchi} can be used to search for SUSY enhancement points within the 5-parameter landscape. At the undeformed origin, the massless gravitini sit in the two first multiplets in eq.~\eqref{eq:CTmultorigin} with ${\rm SO}(3)_{\rm L}\times\ts{{\rm SO}(3)}_{\rm L}\times{\rm SO}(3)_{\rm R}\times\ts{{\rm SO}(3)}_{\rm R}$ spins
\begin{equation}
	\big(\nicefrac{1}{2},\nicefrac{1}{2};0,0\big) \quad {\rm and} \quad \big(0,0;\nicefrac{1}{2},\nicefrac{1}{2}\big)\,,
\end{equation}
and conformal dimension $\Delta=\nicefrac{3}{2}$. Turning on the deformation leads to a split into four groups of two modes, with the following ${\rm U}(1)_{\rm L}\times\ts{{\rm U}(1)}_{\rm L}\times{\rm U}(1)_{\rm R}\times\ts{{\rm U}(1)}_{\rm R}$ charges and conformal dimensions within the 3-dimensional truncation:
\begin{equation}
	\begin{split}
		\left(\pm1,\pm1;0,0\right):\quad \Delta=\frac{1}{2}+\sqrt{1 + \frac{\left(
		\chi_{1}+\chi_{2}+\alpha\,\big(\ts{\chi}_{1}-\ts{\chi}_{2}\big)\right)^{2}}{4\, e^{2\psi}\, (1+\alpha^{2})}}\,, \\
		\left(\pm1,\mp1;0,0\right):\quad \Delta=\frac{1}{2}+\sqrt{1 + \frac{\left(
		\chi_{1}+\chi_{2}-\alpha\,\big(\ts{\chi}_{1}-\ts{\chi}_{2}\big)\right)^{2}}{4\, e^{2\psi}\, (1+\alpha^{2})}}\,, \\
		\left(0,0;\pm1,\pm1\right):\quad \Delta=\frac{1}{2}+\sqrt{1 + \frac{\left(
		\chi_{1}-\chi_{2}+\alpha\,\big(\ts{\chi}_{1}+\ts{\chi}_{2}\big)\right)^{2}}{4\, e^{2\psi}\, (1+\alpha^{2})}}\,, \\
		\left(0,0;\pm1,\mp1\right):\quad \Delta=\frac{1}{2}+\sqrt{1 + \frac{\left(
		\chi_{1}-\chi_{2} - \alpha\,\big(\ts{\chi}_{1}+\ts{\chi}_{2}\big)\right)^{2}}{4\, e^{2\psi}\, (1+\alpha^{2})}}\,.
	\end{split}
\end{equation}
SUSY enhancement points are then given by combinations of the parameters that yield $\Delta = \nicefrac32$. This can either happen by leaving some modes with $\Delta=\nicefrac{3}{2}$ invariant, or where some other, originally massive gravitini outside the consistent truncation, \textit{i.e.} with conformal dimension given by \eqref{eq:confdims3s3s1} with $n \neq 0$, obtain $\Delta = \nicefrac32$. Here we will focus on the supersymmetry enhancement within the 3-dimensional truncation and will discuss the higher KK modes in section \ref{s:Compact}. The different possibilities are listed below.

\begin{itemize}
	\item {\boldmath ${\cal N}=2$} The four-dimensional hypersurfaces
	\begin{equation} \label{eq:N2}
		\chi_{2} = \chi_{1} \pm\alpha\,(\ts{\chi}_{1}+\ts{\chi}_{2})
		\quad {\rm or}\quad \chi_{2} = -\chi_{1} \pm\alpha\,(\ts{\chi}_{1}-\ts{\chi}_{2})
		\,,
	\end{equation}
	give enhancements to ${\cal N}=(0,2)$ and ${\cal N}=(2,0)$, respectively. On this hypersurface \eqref{eq:N2}, the spectrum then organizes into multiplets of $\SU{1\vert1,1}$.
	\item {\boldmath ${\cal N}=4$} Choosing
	\begin{equation} \label{eq:N4Large}
		\begin{cases}
			\chi_{2} = \chi_{1}\,,\\
			\ts{\chi}_{2} = -\ts{\chi}_{1}\,,
		\end{cases}
		\quad {\rm or }\quad
		\begin{cases}
			\chi_{2} = - \chi_{1}\,,\\
			\ts{\chi}_{2} = \ts{\chi}_{1}\,,
		\end{cases}
	\end{equation}
	leads to enhancements to ${\cal N}=(0,4)$ and ${\cal N}=(4,0)$, respectively. On these three-dimensional hyper-surfaces \eqref{eq:N4Large}, the isometry group is $\SO{4}\times {\rm U}(1)^{2}$ and the spectrum reorganizes into multiplets of $D^{1}(2,1;\alpha)$. In the special case
	\begin{equation} \label{eq:limittoT4}
		\begin{cases}
			\chi_{1}=\chi_{2}= 0,\\
			\ts{\chi}_{1}=-\,\ts{\chi}_{2}=\ts{\chi}=\sqrt{1-e^{2\omega}},\\
			\psi = -\,\omega,
		\end{cases}
	\end{equation}
	the solution~\eqref{eq:chipsi} corresponds to the 10-dimensional uplift on $S^{3}\times S^{1}$ of the one-parameter deformation of ${\rm AdS}_{3}\times S^{3}$ preserving ${\cal N}=(0,4)$ supersymmetries of ref.~\cite{Eloy:2021fhc}.
	
	Another possibility is to set
	\begin{equation} \label{eq:N4Small}
		\begin{cases}
			\ts{\chi}_{1} = \pm \chi_{1}/\alpha\,,\\
			\ts{\chi}_{2} = \mp \chi_{2}/\alpha\,,
		\end{cases}
		\quad {\rm or}\quad
		\begin{cases}
			\ts{\chi}_{1} = \pm \chi_{2}/\alpha\,,\\
			\ts{\chi}_{2} = \mp \chi_{1}/\alpha\,,
		\end{cases}
	\end{equation}
	giving 3-parameter families of ${\cal N}=(2,2)$ solutions, with multiplets of $\SU{1\vert1,1}_{\rm L}\times\SU{1\vert1,1}_{\rm R}$.
	\item {\boldmath ${\cal N}=6$} Finally, for
	\begin{equation} \label{eq:N6}
		\begin{cases}
			\chi_{2} = \chi_{1}\,,\\
			\ts{\chi}_{1} = 
			\pm \chi_{1}
			/\alpha\,,\\
			\ts{\chi}_{2} = 
			\mp \chi_{1}
			/\alpha\,,
		\end{cases}
		\quad {\rm or}\quad
		\begin{cases}
			\chi_{2} = -\chi_{1}\,,\\
			\ts{\chi}_{1} = 
			\pm \chi_{1}
			/\alpha\,,\\
			\ts{\chi}_{2} = 
			\mp \chi_{1}
			/\alpha\,,
		\end{cases}
	\end{equation}
	there are enhancements to ${\cal N}=(2,4)$ and ${\cal N}=(4,2)$, respectively. The isometry group is $\SO{4}\times {\rm U}(1)^{2}$ and the spectrum is then given by multiplets of $D^{1}(2,1;\alpha)_{\rm L, R}\times\SU{1\vert1,1}_{\rm R,L}$.
\end{itemize}

In the above cases, it is easy to see from \eqref{eq:confdimshiftchi} that the spectrum  is invariant under $\chi_i \rightarrow \chi_i + 4\pi k_i\,\alpha$, $\ts{\chi}_i \rightarrow \ts{\chi}_i + 4\pi \ts{k}_i$, for $k_i$, $\ts{k}_i \in \mathbb{Z}$, with the SUSY enhancements also occurring if the relations \eqref{eq:N2}, \eqref{eq:N4Large}, \eqref{eq:N4Small} or \eqref{eq:N6} are satisfied up to integer shifts. We will return to this point below in section \ref{s:Compact}, when discussing compactness of the moduli space.

For generic values of $\chi_i$, $\ts{\chi}_i$, the AdS$_3$ vacuum will be completely non-supersymmetric. However, it is easy to see from inspection of the spectrum \eqref{eq:confdims3s3s1} and \eqref{eq:confdimshiftchi} that the spectrum is bounded from below by the masses of the 3-dimensional gauged supergravity modes at the ${\cal N}=(4,4)$ supersymmetric origin, \textit{i.e.} where $\chi_i = \ts{\chi}_i = n = \ell = \ts{\ell} = 0$. Therefore, these non-supersymmetric AdS$_3$ vacua are perturbatively stable within IIB supergravity and, similarly to \cite{Giambrone:2021wsm} suggest that there may be a dual non-supersymmetric conformal manifold if there are no instabilities beyond supergravity. However, just as in \cite{Giambrone:2021wsm}, the non-SUSY vacua are protected against perturbative $\alpha'$ or $g_s$ corrections of string theory, which might have lifted the moduli $\chi_i$, $\ts{\chi}_i$, and there are no scalars at the BF bound which may have been pushed below by such corrections. The moduli $\chi_i$, $\ts{\chi}_i$ locally correspond to diffeomorphisms and, therefore, any diffeomorphism-invariant quantity, such as higher powers of the curvature tensor or fluxes, as would appear in $\alpha'$ or $g_s$ corrections, will be independent of $\chi_i$ and $\ts{\chi}_i$. The vacua are also protected against Witten's bubbles of nothing \cite{Witten:1981gj}, due to flux quantisation on $S^3$ and the spin structure on $S^1$. This does not rule out all potential sources of instabilities. For example, the non-SUSY vacua may suffer from non-perturbative instabilities triggered by brane nucleation \cite{Ooguri:2016pdq,Bomans:2021ara} or non-vanishing $\beta$-functions of marginal multi-trace operators~\cite{Witten:2001ua}. However, our non-SUSY vacua here may have further protection against such instabilities due to their continuous limits to the SUSY vacua, just as they have against perturbative higher-derivative corrections.

\paragraph{\boldmath $(\zeta,\ts{\zeta},\Omega,\ts{\Omega},\psi)$}
Restricting to the parameters $(\zeta_i, \ts{\zeta}_i, \Omega, \ts{\Omega}, \psi)$, $i = 1,2$, and using again Hopf coordinates, the metric is
\begin{equation} \label{eq:metriczeta}
	\begin{split}
		\d s^{2} &= \ell_{\rm AdS}^{2}\,\d s^{2}\left({\rm AdS}_{3}\right) -\Delta\Big(\cos^{2}(\theta)+e^{\Omega}\sin^{2}(\theta)\Big)\Big(e^{\ts{\Omega}}\cos^{2}(\ts{\theta})+\sin^{2}(\ts{\theta})\Big)\,\d z^{2}\\
		&+ \alpha^{2}\Big[\d\theta^{2} + \Delta\Big(e^{\ts{\Omega}}\cos^{2}(\ts{\theta})+\sin^{2}(\ts{\theta})\Big)\Big(e^{\Omega}\cos^{2}(\theta)\,\d\phi_{1}^{2}+\sin^{2}(\theta)\,\d\phi_{2}^{2}\Big)\Big]\\
		&+\d\ts{\theta}^{2} + \Delta\Big(\cos^{2}(\theta)+e^{\Omega}\sin^{2}(\theta)\Big)\Big(\cos^{2}(\ts{\theta})\,\d\ts{\phi}_{1}^{2}+e^{\ts{\Omega}}\sin^{2}(\ts{\theta})\,\d\ts{\phi}_{2}^{2}\Big)\\
		&+\Delta\,e^{2\psi}\,e^{\Omega}\cos^{2}(\theta)\cos^{2}(\ts{\theta}) \left(\alpha\,\ts{\zeta}_{2}\,\d\phi_{1}-\zeta_{2}\,\d\ts{\phi}_{1}\right)^{2}+\Delta\,e^{2\psi}\,e^{\ts{\Omega}}\sin^{2}(\theta)\sin^{2}(\ts{\theta}) \left(\alpha\,\ts{\zeta}_{1}\,\d\phi_{2}-\zeta_{1}\,\d\ts{\phi}_{2}\right)^{2}\\
		&+\Delta\,e^{2\psi}\,e^{\Omega+\ts{\Omega}}\cos^{2}(\theta)\sin^{2}(\ts{\theta}) \left(\alpha\,\ts{\zeta}_{1}\,\d\phi_{1}-\zeta_{2}\,\d\ts{\phi}_{2}\right)^{2}+\Delta\,e^{2\psi}\,\sin^{2}(\theta)\cos^{2}(\ts{\theta}) \left(\alpha\,\ts{\zeta}_{2}\,\d\phi_{2}-\zeta_{1}\,\d\ts{\phi}_{1}\right)^{2}\\
		&-2\,\alpha\,\Delta\,e^{2\psi}\left(\zeta_{1}\,\cos^{2}(\theta)\,\d\phi_{1}+e^{\Omega}\,\zeta_{2}\,\sin^{2}(\theta)\,\d\phi_{2}\right)\left(e^{\ts{\Omega}}\,\ts{\zeta}_{1}\,\cos^{2}(\ts{\theta})\,\d\ts{\phi}_{1}+\ts{\zeta}_{2}\,\sin^{2}(\ts{\theta})\,\d\ts{\phi}_{2}\right)\\
		&+\Delta\,e^{2\psi}\Big(e^{\ts{\Omega}}\cos^{2}(\ts{\theta})+\sin^{2}(\ts{\theta})\Big)\Big[\cos^{2}(\theta)\left(\alpha\,\zeta_{1}\,\d\phi_{1}+\d z\right)^{2}+e^{\Omega}\sin^{2}(\theta)\left(\alpha\,\zeta_{2}\,\d\phi_{2}+\d z\right)^{2}\Big]\\
		&+\Delta\,e^{2\psi}\Big(\cos^{2}(\theta)+e^{\Omega}\sin^{2}(\theta)\Big)\Big[e^{\ts{\Omega}}\cos^{2}(\ts{\theta})\left(\ts{\zeta}_{1}\,\d\ts{\phi}_{1}-\d z\right)^{2}+\sin^{2}(\ts{\theta})\left(\ts{\zeta}_{2}\,\d\ts{\phi}_{2}-\d z\right)^{2}\Big],
	\end{split}
\end{equation}
with the warp factor
\begin{equation}
	\begin{aligned}
	\Delta^{-1} &= \Big(\cos^{2}(\theta)+e^{\Omega}\sin^{2}(\theta)\Big)\Big(e^{\ts{\Omega}}\cos^{2}(\ts{\theta})+\sin^{2}(\ts{\theta})\Big)\\
	&+e^{2\psi}\Big(\cos^{2}(\theta)+e^{\Omega}\sin^{2}(\theta)\Big)\left(\ts{\zeta}_{2}^{2}\cos^{2}(\ts{\theta})+e^{\ts{\Omega}}\,\ts{\zeta}_{1}^{2}\sin^{2}(\ts{\theta})\right)\\
	&+e^{2\psi}\Big(e^{\ts{\Omega}}\cos^{2}(\ts{\theta})+\sin^{2}(\ts{\theta})\Big)\left(e^{\Omega}\,\zeta_{2}^{2}\cos^{2}(\theta)+\zeta_{1}^{2}\sin^{2}(\theta)\right).
	\end{aligned}
\end{equation}
To provide an expression for $H_{(3)}$ it is convenient to first define some forms:
\begin{equation}
  \begin{aligned}
    \Phi_{1} &= \alpha\,\Big(\big(e^{\ts{\Omega}}\cos^{2}(\ts{\theta})+\sin^{2}(\ts{\theta})\big)\big(e^{\Omega}+e^{2\psi}\,\zeta_{1}^{2}\big)+e^{2\psi+\Omega}\left(\ts{\zeta}_{2}^{2}\cos^{2}(\ts{\theta})+e^{\ts{\Omega}}\,\ts{\zeta}_{1}^{2}\sin^{2}(\ts{\theta})\right)\Big)\,\d\phi_{1},\\
    \Phi_{2} &= \alpha\,\Big(\big(e^{\ts{\Omega}}\cos^{2}(\ts{\theta})+\sin^{2}(\ts{\theta})\big)\big(1+e^{2\psi+\Omega}\,\zeta_{2}^{2}\big)+e^{2\psi}\left(\ts{\zeta}_{2}^{2}\cos^{2}(\ts{\theta})+e^{\ts{\Omega}}\,\ts{\zeta}_{1}^{2}\sin^{2}(\ts{\theta})\right)\Big)\,\d\phi_{2},\\
    \ts{\Phi}_{1} &= \Big(\big(\cos^{2}(\theta)+e^{\Omega}\sin^{2}(\theta)\big)\big(1+e^{2\psi+\ts{\Omega}}\,\ts{\zeta}_{1}^{2}\big)+e^{2\psi}\left(e^{\Omega}\,\zeta_{2}^{2}\cos^{2}(\theta)+\zeta_{1}^{2}\sin^{2}(\theta)\right)\Big)\,\d\ts{\phi}_{1},\\
    \ts{\Phi}_{2} &= \Big(\big(\cos^{2}(\theta)+e^{\Omega}\sin^{2}(\theta)\big)\big(e^{\ts{\Omega}}+e^{2\psi}\,\ts{\zeta}_{2}^{2}\big)+e^{2\psi+\ts{\Omega}}\left(e^{\Omega}\,\zeta_{2}^{2}\cos^{2}(\theta)+\zeta_{1}^{2}\sin^{2}(\theta)\right)\Big)\,\d\ts{\phi}_{2},\\
    \varphi_{1} &= \alpha\,e^{2\psi}\cos^{2}(\theta)\,\d\phi_{1},\qquad\qquad\qquad
    \varphi_{2} = \alpha\,e^{2\psi}\sin^{2}(\theta)\,\d\phi_{2},\\
    \ts{\varphi}_{1} &= e^{2\psi}\cos^{2}(\ts{\theta})\,\d\ts{\phi}_{1},\qquad\qquad\qquad\ \ \,
    \ts{\varphi}_{2} = e^{2\psi}\sin^{2}(\ts{\theta})\,\d\ts{\phi}_{2}.
  \end{aligned}
\end{equation}
In terms of these forms, we get
\begin{equation}
  \begin{aligned}
    H_{(3)} &= 2\,\ell_{\rm AdS}^{2}\,{\rm Vol}({\rm AdS_{3}})\\
    & + 2\,\Delta^{2}\cos(\theta)\sin(\theta)\,\d\theta\wedge \bigg[\Phi_{1}\wedge\Phi_{2} + \left(\zeta_{1}\ts{\zeta}_{2}+e^{\Omega+\ts{\Omega}}\zeta_{2}\ts{\zeta}_{1}\right)\Big(\Phi_{1}\wedge\ts{\varphi}_{1}-\Phi_{2}\wedge\ts{\varphi}_{2}\Big)\\
    &\quad+ \left(e^{\ts{\Omega}}\zeta_{1}\ts{\zeta}_{1}+e^{\Omega}\zeta_{2}\ts{\zeta}_{2}\right)\Big(\Phi_{1}\wedge\ts{\varphi}_{2}-\Phi_{2}\wedge\ts{\varphi}_{1}\Big)+\Big(\zeta_{1}^{2}-e^{2\Omega}\zeta_{2}^{2}\Big)\Big(e^{2\ts{\Omega}}\ts{\zeta}_{1}^{2}-\ts{\zeta}_{2}^{2}\Big)\,\ts{\varphi}_{1}\wedge\ts{\varphi}_{2}\\
    &\quad+e^{2\psi}\Big(e^{\ts{\Omega}}\cos^{2}(\ts{\theta})+\sin^{2}(\ts{\theta})\Big)\bigg(\Big(e^{\Omega}\zeta_{2}\,\Phi_{1}-\zeta_{1}\,\Phi_{2}\Big)-\Big(\zeta_{1}^{2}-e^{2\Omega}\zeta_{2}^{2}\Big)\Big(\ts{\zeta}_{2}\,\ts{\varphi}_{1}+e^{\ts{\Omega}}\ts{\zeta}_{1}\,\ts{\varphi}_{2}\Big)\bigg)\wedge\d z\bigg]\\
    & + 2\,\Delta^{2}\cos(\ts{\theta})\sin(\ts{\theta})\,\d\ts{\theta}\wedge \bigg[\ts{\Phi}_{1}\wedge\ts{\Phi}_{2} + \left(\zeta_{1}\ts{\zeta}_{2}+e^{\Omega+\ts{\Omega}}\zeta_{2}\ts{\zeta}_{1}\right)\Big(\ts{\Phi}_{1}\wedge\varphi_{1}-\ts{\Phi}_{2}\wedge\varphi_{2}\Big)\\
    &\quad+ \left(e^{\ts{\Omega}}\zeta_{1}\ts{\zeta}_{1}+e^{\Omega}\zeta_{2}\ts{\zeta}_{2}\right)\Big(\ts{\Phi}_{1}\wedge\varphi_{2}-\ts{\Phi}_{2}\wedge\varphi_{1}\Big)+\Big(\zeta_{1}^{2}-e^{2\Omega}\zeta_{2}^{2}\Big)\Big(e^{2\ts{\Omega}}\ts{\zeta}_{1}^{2}-\ts{\zeta}_{2}^{2}\Big)\,\varphi_{1}\wedge\varphi_{2}\\
    &\quad+e^{2\psi}\Big(\cos^{2}(\theta)+e^{\Omega}\sin^{2}(\theta)\Big)\bigg(\Big(\ts{\zeta}_{2}\,\ts{\Phi}_{1}-e^{\ts{\Omega}}\ts{\zeta}_{1}\,\ts{\Phi}_{2}\Big)-\Big(e^{2\ts{\Omega}}\ts{\zeta}_{1}^{2}-\ts{\zeta}_{2}^{2}\Big)\Big(e^{\Omega}\zeta_{2}\,\varphi_{1}+\zeta_{1}\,\varphi_{2}\Big)\bigg)\wedge\d z\bigg]. 
  \end{aligned}
\end{equation}
Finally, the metric parametrising this moduli space is:
\begin{equation}
	\begin{aligned}
		\d s^{2}_{\rm mod.} &= -\frac{1}{2}\bigg[e^{2\psi}\left(e^{-\Omega}\,\d\zeta_{1}^{2}+e^{\Omega}\,\d\zeta_{2}^{2}+e^{\ts{\Omega}}\,\d\ts{\zeta}_{1}^{2}+e^{-\ts{\Omega}}\,\d\ts{\zeta}_{2}^{2}\right) +2\,\d\psi^{2} + \frac{1}{2}\,\d\Omega^{2}+ \frac{1}{2}\,\d\ts{\Omega}^{2}\bigg]\,.
	\end{aligned}
\end{equation}
As previously, the deformation breaks the isometry group down to ${\rm U}(1)_{\rm L}\times{\rm U}(1)_{\rm R}\times\ts{{\rm U}(1)}_{\rm L}\times\ts{{\rm U}(1)}_{\rm R}$ and all supersymmetries. Note that for
\begin{equation}
	\zeta_{1}=\pm\zeta_{2},\quad \ts{\zeta}_{1}=\mp\ts{\zeta}_{2},\quad\psi=\Omega=\ts{\Omega}=0,
\end{equation}
the solution~\eqref{eq:metriczeta} reproduces eq.~\eqref{eq:chipsi} for
\begin{equation}
	\chi_{1}=\pm\chi_{2}=\frac{\zeta_{1}}{1+\zeta_{1}^{2}+\ts{\zeta}_{1}^{2}},\quad \ts{\chi}_{1}=\mp\ts{\chi}_{2}=-\frac{\ts{\zeta}_{1}}{1+\zeta_{1}^{2}+\ts{\zeta}_{1}^{2}},\quad e^{2\psi}=\frac{1}{(1+\zeta_{1}^{2}+\ts{\zeta}_{1}^{2})^{2}},
\end{equation}
with ${\cal N}=(4,0)$ (or $(0,4)$) and symmetry enhancement to ${\rm SO}(4)\times{\rm U}(1)^{2}$. There seems to be no other points with (super)-symmetry enhancement in the scalar manifold.

\subsection{Compactification of the moduli space} \label{s:Compact}

We can use the Kaluza-Klein spectrum and the uplift of the 3-dimensional gauged supergravity models to IIB supergravity to understand properties of the dual CFTs that are obscured in the lower-dimensional supergravities. One such aspect is the compactness of the conformal manifold. Consider the deformations $\chi_i$, $\ts{\chi}_i$, which appear non-compact in 3-dimensional ${\cal N}=16$ gauged supergravity, with only the point $\chi_i = \ts{\chi}_i = 0$ having ${\cal N}=(4,4)$ supersymmetry. The subset of these deformations corresponding to the limit~\eqref{eq:limittoT4} was analysed in the half-maximal theory in ref.~\cite{Eloy:2021fhc}. Again, within 3-dimensional ${\cal N}=8$ supergravity, this deformation is non-compact, and remains non-compact even after uplifting to 6-dimensional supergravity on $S^3$, with no periodicity in $\ts{\chi}$ observed in the KK spectrum of the 6-dimensional solution \cite{Eloy:2021fhc}.

However, from  \eqref{eq:spectrumorigin}, \eqref{eq:confdims3s3s1}, \eqref{eq:confdimshiftchi} we see that the full Kaluza-Klein spectrum is invariant under the shifts
\begin{equation} \label{eq:ShiftSymmetries}
	\begin{split}
		\chi_i &\rightarrow \chi_i + 4\pi\alpha\, k_i\,, \\
		\ts{\chi}_i &\rightarrow \ts{\chi}_i + 4\pi \ts{k}_i \,,
	\end{split}
\end{equation}
for $k_i$, $\ts{k}_i \in \mathbb{Z}$. In these cases, higher KK modes on the $S^1$ outside the ${\cal N}=16$ truncation replace those within the consistent truncation. Therefore, even though neither the 3-dimensional gauged supergravity, nor 6-dimensional supergravity is invariant under these shifts, upon lifting to the full 10-dimensional IIB theory, the higher KK modes on $S^1$ restore the invariance under the shifts, just as in the analogous deformations studied in AdS$_4 \times S^5 \times S^1$ \cite{Giambrone:2021zvp,Giambrone:2021wsm}. This highlights the importance of the $S^1$ direction in studying these AdS$_3$ vacua.

The compactification of the moduli space in 10 dimensions can also be understood geometrically. Each of $\chi_i$, $\ts{\chi}_i$ corresponds to the real part of the complex structure modulus of a $T^2$ fibration with local coordinates $\phi_i$, $\ts{\phi}_i$ together with $z$, respectively. Explicitly, the complex structures for the $T^2_{(\phi_i,\, z)}$ and $T^2_{(\ts{\phi}_i,\, z)}$ are given by
\begin{equation} \label{eq:T2CStructure}
	\tau_i = i\,\frac{e^\psi}{\alpha} + \frac1\alpha \chi_i \,, \qquad \ts{\tau}_i = i\,e^\psi + \ts{\chi}_i \,.
\end{equation}
Under this identification, the shifts \eqref{eq:ShiftSymmetries} correspond to $T$-transformations of the tori, which belong to the modular group $\SL{2,\mathbb{Z}}$, just as for AdS$_4 \times S^5 \times S^1$ \cite{Giambrone:2021zvp,Giambrone:2021wsm}. In fact, \eqref{eq:T2CStructure} only shows how the bosonic fields transform and thus suggests that the periodicities are $2\pi\alpha$ and $2\pi$. However, this is wrong: as the KK spectrum \eqref{eq:confdimshiftchi} shows, the fermions require double that periodicity, \textit{i.e.} $4\pi\alpha$ and $4\pi$ as in \eqref{eq:ShiftSymmetries}.

Given the interpretation of the shifts \eqref{eq:ShiftSymmetries} as $T$-transformations of a $T^2$-fibration, it is natural to ask whether $S$-transformations of these $T^2$ would also leave the AdS$_3 \times S^3 \times S^3 \times S^1$ solution invariant. The $S$-transformation acts on the moduli as
\begin{equation} \label{eq:Stransformchi}
	\chi_i \rightarrow - \frac{\alpha^2 \chi_i}{\chi_i^2+e^{2\psi}} \,, \qquad e^{\psi} \rightarrow \frac{\alpha^2\, e^{\psi}}{\chi_i^2+e^{2\psi}} \,,
\end{equation}
for the $T^2_{(\phi_i,\, z)}$ and
\begin{equation} \label{eq:Stransformchitilde}
	\ts{\chi}_i \rightarrow - \frac{\ts{\chi}_i}{\ts{\chi}_i^2+e^{2\psi}} \,, \qquad e^{\psi} \rightarrow \frac{e^{\psi}}{\ts{\chi}_i^2+e^{2\psi}} \,,
\end{equation}
for the $T^2_{(\ts{\phi}_i,\, z)}$. Thus, the $S$-transformation effectively inverts the radius of the $S^1$, while transforming the combinations
\begin{equation} \label{eq:STransformChi}
	e^{-\psi}\chi_i \rightarrow - e^{-\psi}\chi_i \,, \qquad e^{-\psi} \ts{\chi}_i \rightarrow - e^{-\psi}\ts{\chi}_i \,.
\end{equation}
If we consider a truncation where the $n \neq 0$ modes are discarded, for example, by uplifting only on $S^3$ to 6 dimensions as in \cite{Eloy:2021fhc} or on $S^3 \times S^3 \times S^1$ to 9 dimensions, then the $S$-transformation clearly leaves the KK spectrum \eqref{eq:spectrumorigin}, \eqref{eq:confdims3s3s1}, \eqref{eq:confdimshiftchi} invariant. This is because the spectrum for the $n = 0$ modes only depends on the combinations $e^{-\psi}\chi_i$ and $e^{-\psi}\ts{\chi}_i$, which pick up minus signs, as in \eqref{eq:STransformChi}. The effect of \eqref{eq:Stransformchi} is to simply exchanges modes of the opposite charges under the ${\rm U}(1)_{\rm L}\times\ts{{\rm U}(1)}_{\rm L}\times{\rm U}(1)_{\rm R}\times\ts{{\rm U}(1)}_{\rm R}$ within the spectrum.

However, if we include the $n \neq 0$ modes by going to the full 10-dimensional IIB supergravity theory, the KK spectrum \eqref{eq:spectrumorigin}, \eqref{eq:confdims3s3s1}, \eqref{eq:confdimshiftchi} is not invariant under the $S$ transformation. At the same time, the $S$-transformation effectively inverts the radius of the $S^1$, as in \eqref{eq:STransformChi} and, thus, takes us out of the regime of validity of supergravity, with new string degrees of freedom potentially becoming light. Indeed, in this case, after the transformation \eqref{eq:STransformChi}, strings wound around $S^1$ will become light, and one would expect them to replace the KK modes that have become heavy, realising the T-duality on $S^1$. This suggests that in the full IIB string theory, the moduli space of each $T^2$ fibration with complex structure \eqref{eq:T2CStructure} should be identified under the M\"{o}bius group $\SL{2,\mathbb{Z}}$, which has a natural action on the metric of this 5-parameter submanifold of moduli space~\eqref{eq:Zam7Param}. Note that this is quite different from the AdS$_4 \times S^5 \times S^1$ ``S-fold'' vacua \cite{Giambrone:2021zvp,Giambrone:2021wsm}, which do not have a $S^1$ rescaling modulus and, therefore, do not carry the natural action of $\SL{2,\mathbb{Z}}$ on each complex structure modulus that we find here.

\section{Conclusions} \label{s:Conclusions}
In this paper, we used $\EE$ ExFT to construct the consistent truncation of IIB supergravity around the ${\cal N}=(4,4)$ AdS$_3 \times S^3 \times S^3 \times S^1$ vacuum. Even though the vacuum breaks half of the supersymmetries, we are able to construct a consistent truncation to 3-dimensional maximal gauged supergravity, providing an uplift to the ${\cal N}=16$ gauged supergravity proposed in \cite{Hohm:2005ui} for the 3-dimensional description of this AdS$_3$ vacuum. We constructed our consistent truncation by doing a $S^1$ reduction of the truncation of IIB on $S^3 \times S^3$ \cite{Inverso:2016eet}, and turning on an additional $H_{(7)}$ flux on $S^3 \times S^3 \times S^1$, which is necessary to obtain an AdS$_3$ vacuum. We also show that it is not possible to turn on such an additional flux for the other ``dyonic'' consistent truncations on $S^6$, $S^5 \times S^1$, $S^4 \times S^2$ \cite{Guarino:2015jca,Inverso:2016eet} reduced on $S^1$, whilst maintaining a consistent truncation to ${\cal N}=16$ gauged supergravity.

To find this vacuum, we had to construct the potential of 3-dimensional ${\cal N}=16$ gauged supergravity in terms of the embedding tensor, which was not previously known. Since the vector fields of ${\cal N}=16$ gauged supergravity transform in the adjoint of $\EE$, the Cartan-Killing metric of $\EE$ can be used to construct new terms in the potential, which leads to a different form of the potential compared to the universal one found in $D \geq 4$ dimensions. Nonetheless, we used the generalised Scherk-Schwarz Ansatz of $\EE$ ExFT \cite{Galli:2022idq} to obtain the potential in terms of the embedding tensor, for all gauged supergravities that can be uplifted to maximal 10-/11-dimensional supergravity. This condition of being able to uplift imposes additional quadratic constraints on the embedding tensor~\eqref{eq:reltrX}, beyond the usual quadratic constraints of gauged supergravity. This can be used to rule out the higher-dimensional origin of some 3-dimensional ${\cal N}=16$ gauged supergravities, as we illustrated in appendix~\ref{app:rulingoutgauging} on one ${\cal N}=(8,0)$ ${\rm AdS}_{3}$ vacuum of ref.~\cite{Deger:2019tem}.

Using our consistent truncation and the 3-dimensional potential, we studied the moduli space of the AdS$_3$ vacuum within 3-dimensional gauged supergravity. Even a simple search led to a surprisingly large 11-dimensional moduli space of AdS$_3 \times S^3 \times S^3 \times S^1$ vacua, with more moduli likely. While most of these moduli break all supersymmetries, we also identified submanifolds where ${\cal N}=(4,2),\, (4,0),\, (2,2),\, (2,0)$ is preserved, where the ${\cal N}=4$ cases always correspond to the ``large'' superconformal algebra. The corresponding moduli are analogous to the  ``flat deformations'' that arise in the AdS$_4 \times S^5 \times S^1$ vacuum \cite{Guarino:2020gfe,Guarino:2021hrc,Giambrone:2021zvp,Giambrone:2021wsm}. \textit{i.e.} they correspond to local diffeomorphisms mixing angles on the $S^3$'s with the $S^1$, but which are not globally well-defined, much like complex structure deformations of $T^2$, allowing us to compute the full Kaluza-Klein spectrum of these deformed AdS$_3$ vacua. Generic ``flat deformations'' of this type break all supersymmetries but are still perturbatively stable. Moreover, they are protected against perturbative higher-derivative corrections because the deformations locally correspond to diffeomorphisms.

We also saw that uplifting our AdS$_3$ vacua to IIB string theory leads to the compactness of some directions of moduli space. The ``flat deformations'' that we studied appear non-compact within the 3-dimensional gauged supergravity, but upon uplifting to 10 dimensions, the moduli become compact, similar to \cite{Giambrone:2021zvp,Giambrone:2021wsm}. Interestingly, this compactness is not visible upon only uplifting on $S^3$ to 6-dimensional supergravity \cite{Eloy:2021fhc}, but require the uplift on the additional $S^1$ direction. The compactification of the moduli can be understood geometrically as the $T$-action of $\SL{2,\mathbb{Z}}$ on $T^2$-fibration. Differently from the AdS$_4 \times S^5 \times S^1$ case \cite{Guarino:2020gfe,Guarino:2021hrc,Giambrone:2021zvp,Giambrone:2021wsm}, in the AdS$_3 \times S^3 \times S^3 \times S^1$ case we also have a $S^1$ rescaling modulus, so that we naturally have the full complex structure moduli related to the $T^2$-fibrations and which carry a natural action of the full $\SL{2,\mathbb{Z}}$. While the $S$-action of $\SL{2,\mathbb{Z}}$ left the Kaluza-Klein spectrum of 6-dimensional supergravity on AdS$_3 \times S^3$ \cite{Eloy:2021fhc} invariant, it is broken by the additional $S^1$ direction when uplifting to IIB supergravity. However, we argued that the $S$-action leave invariant the full AdS$_3 \times S^3 \times S^3 \times S^1$ solution of IIB string theory, once string winding modes around $S^1$ are included. In this case, the complex structure deformations live in the fundamental domain of $\SL{2,\mathbb{Z}}$, rather than the whole $\mathbb{C}$-plane, as the 3-dimensional supergravity would suggest. 

Our work raises several questions that deserve further inquiry. Firstly, we only performed a simple search, yielding an 11-dimensional moduli space of AdS$_3 \times S^3 \times S^3 \times S^1$ within the ${\cal N}=16$ gauged supergravity. It would be good to complete our analysis to compute the full 3-dimensional moduli space and to understand why it is so large. Our consistent truncation Ansatz could then be used to uplift the moduli space to IIB supergravity, and to use Kaluza-Klein spectrometry \cite{Malek:2019eaz,Malek:2020yue,Eloy:2020uix} to obtain the conformal dimensions of all single-trace operators and determine the perturbative stability of the non-SUSY vacua. This would shed light on the dual conformal manifold. Secondly, it would be interesting to study the string sigma-model \cite{Eberhardt:2019niq} on the ``flat deformations'' of AdS$_3 \times S^3 \times S^3 \times S^1$, or the analogous deformations of AdS$_3 \times S^3 \times T^4$. Since these deformations locally are diffeomorphisms, one may expect a simple worldsheet description on the ${\cal N}=(4,2),\, (4,0),\, (2,2),\, (2,0)$ SUSY vacua and, potentially, even for the non-SUSY AdS$_3 \times S^3 \times S^3 \times S^1$ vacua obtained this way. The non-perturbative stability of the perturbatively stable non-supersymmetric AdS$_3$ vacua we found also deserves further study. Thirdly, similar ``flat deformations'' arise in any string compactification involving an $S^1$, in particular AdS$_3 \times S^3 \times T^4$. It would be interesting to understand universal features of these deformations and their CFT duals. Finally, we would like to use the tools developed here to study RG flows between solutions of three-dimensional supergravity, or equivalently between two-dimensional conformal field theories. Note however that this cannot be applied to the RG flow of ref.~\cite{Berg:2001ty}: their theory cannot be described using the generalized Scherk-Schwarz Ansatz of \EE or $\SO{8,8}$ exceptional field theory as reproducing their embedding tensor would require breaking the section constraint. We leave these questions for future work.

\section*{Acknowledgements}
We are grateful to Nikolay Bobev, Adolfo Guarino, Gabriel Larios, Yolanda Lozano, Niall Macpherson, Michela Petrini, Colin Sterckx and Dan Waldram for useful discussions and correspondence. CE is supported by the FWO-Vlaanderen through the project G006119N and by the Vrije Universiteit Brussel through the Strategic Research Program ``High-Energy Physics''. MG and EM are supported by the Deutsche Forschungsgemeinschaft (DFG, German Research Foundation) via the Emmy Noether program ``Exploring the landscape of string theory flux vacua using exceptional field theory'' (project number 426510644).

\appendix

\section{Ruling out gaugings} \label{app:rulingoutgauging}
As mentioned in section~\ref{s:Potential}, eq.~\eqref{eq:reltrX} can be used to assess whether a given three-dimensional theory of embedding tensor $X_{\fl{MN}}$ has a higher-dimensional origin. Let us illustrate that on the ${\cal N}=(8,0)$ ${\rm AdS}_{3}$ vacuum with $\GL{6} \times \SO{2,2}$ gauging constructed in ref.~\cite{Deger:2019tem}. This theory has a compact gauging smaller than $\SO{9}$, thus escaping the no-go result of ref.~\cite{Galli:2022idq}. The embedding tensor of the theory is expressed by breaking $\EE$ as
\begin{equation}
	\begin{aligned}
		\EE &\longrightarrow \SO{8,8}\\
		\mathbf{248} &\longrightarrow \mathbf{120} \oplus \mathbf{128}\\
		X^{\fl{M}} &\longrightarrow \left\{X^{[\fl{\sf IJ}]},X^{\fl{\sf A}}\right\}\,,
	\end{aligned}
\end{equation}
with $\fl{\sf I}$ and $\fl{\sf A}$ labels of the fundamental and spinorial representations of $\SO{8,8}$, respectively. Explicitly,
\begin{equation}
	\begin{aligned}
		X_{\fl{\sf IJ},\fl{\sf KL}} &= \Theta_{\fl{\sf IJ},\fl{\sf KL}}\,,\\
		X_{\fl{\sf AB}} &= -2\,\theta\,\eta_{\fl{\sf AB}}+\frac{1}{24}\,\Gamma^{\fl{\sf IJ},\fl{\sf KL}}_{\fl{\sf AB}}\,\Theta_{\fl{\sf IJKL}}\,,\\
		X_{\fl{\sf IJ},\fl{\sf A}} &=0\,,
	\end{aligned}
\end{equation}
where $\Gamma^{\fl{\sf IJKL}}_{\fl{\sf AB}}$ is the four-fold product of $\SO{8,8}$ $\Gamma$ matrices and $\eta_{\fl{\sf AB}}$ the $\mathbf{128}\otimes\mathbf{128}$ component of the Cartan-Killing metric, or equivalently the $\SO{8,8}$ charge conjugation matrix. The tensor $\Theta_{\fl{\sf IJ},\fl{\sf KL}}$ is given by
\begin{equation}
	\Theta_{\fl{\sf IJ},\fl{\sf KL}}=\theta_{\fl{\sf IJKL}}+\frac12\big(\eta_{\fl{\sf K}[\fl{\sf I}}\theta_{\fl{\sf J}]\fl{\sf L}}-\eta_{\fl{\sf L}[\fl{\sf I}}\theta_{\fl{\sf J}]\fl{\sf K}}\big)+\theta\,\eta_{\fl{\sf K}[\fl{\sf I}}\eta_{\fl{\sf J}]\fl{\sf L}}\,,
\end{equation}
with $\theta_{\fl{\sf IJKL}}$ totally antisymmetric and $\theta_{\fl{\sf IJ}}$ symmetric. They have the following non-vanishing components:
\begin{align}
  & \theta_{ij}=\delta_{ij}\,,\quad \theta_{\alpha\beta}=-3\,\delta_{\alpha\beta}\,,\quad \theta_{r_{+}s_{+}}=-\delta_{r_{+}s_{+}}\,,\quad \theta_{r_{-}s_{-}}=3\,\delta_{r_{-}s_{-}}\,,\nonumber\\
  & \theta=-1\,,\quad \theta_{\alpha\beta r_{-}s_{-}} = -3\,\epsilon_{\alpha\beta}\,\epsilon_{r_{-}s_{-}}\,,\quad \theta_{ij r_{+}s_{+}} = 2\,\delta_{i[r_{+}}\,\delta_{s_{+}]j}\,,
\end{align}
where we split the indices according to
\begin{equation}
	\begin{aligned}
		\SO{8,8} & \longrightarrow \SO{8}\times\SO{8} \longrightarrow \SO{6}\times\SO{2}\times\SO{6}\times\SO{2} \\
		X^{\fl{\sf I}} & \ \tikz{\draw[arrows={->[scale=1.1]}] (0,0) -- (3.55
	,0);}\ \left\{X^{i},\,X^{\alpha},\,X^{r_{+}},\,X^{r_{-}}\right\}\,,
	\end{aligned}
\end{equation}
and defined the matrix
\begin{equation}
	 \epsilon = \begin{pmatrix}
                0 & 1\\
                -1 & 0
              \end{pmatrix}\,.
\end{equation}
With this embedding, we get
\begin{equation}
	X_{\ov{MN}}\,X^{\ov{MN}} = \frac{55}{1922}\left(X_{\ov M}{}^{\ov M}\right)^2\,,
\end{equation}
thus violating the condition~\eqref{eq:reltrX} (the trombone gauging vanishes for the truncation). Then, we deduce that the ${\cal N}=(8,0)$ theory of ref.~\cite{Deger:2019tem} with gauging $\GL{6} \times \SO{2,2}$ does not admit a higher-dimensional origin as constructed with \EE exceptional field theory. This illustrates how the constraint~\eqref{eq:reltrX} can be used to determine if a given three-dimensional gauged supergravity can be uplifted.

\section{The $S^n$ twist matrix} \label{s:SLn}
Let us construct the $\SL{4} \simeq \SO{3,3}$ twist matrix that describes the consistent truncation on $S^3$. This can be viewed as a special case of a family of consistent truncations on $S^n$ \cite{Lee:2014mla,Hohm:2014qga,Baguet:2015iou}. Here, we give a slightly different, but equivalent, form of this construction.

We use an $\SL{n+1}$ ExFT, like in \cite{Lee:2014mla}, which encodes an $n$-dimensional metric and volume-form flux, \textit{i.e.} Freund-Rubin compactifications, and thus gives a natural description of $S^n$ compactifications. The $\SL{n+1}$ ExFT formally has coordinates in the antisymmetric representation of $\SL{n+1}$, $y^{IJ} = - y^{JI}$, with $I = 1, \ldots, n+1$, and similarly, generalised vector fields transform in the antisymmetric representation of $\SL{n+1}$. The generalised Lie derivative on a generalised tensor in the fundamental $V^I$ of weight $\lambda$ is given by
\begin{equation} \label{eq:SLnGenLie}
	\gL_\Lambda V^I = \frac12 \Lambda^{JK} \partial_{JK} V^{I} - V^J \partial_{JK} \Lambda^{IK} + \left( \frac{\lambda}{2} + \frac{1}{n+1} \right) V^I \partial_{JK} \Lambda^{JK} \,,
\end{equation}
and similarly for other generalised tensors. Closure of the generalised Lie derivative \eqref{eq:SLnGenLie} requires the section condition
\begin{equation} \label{eq:SLnSC}
	\partial_{[IJ} \otimes \partial_{KL]} = 0 \,,
\end{equation}
which restricts the dependence of all fields to a subset of physical coordinates.

We are interested in the maximal solutions of the section condition \eqref{eq:SLnSC} which preserve a $\SL{n} \subset \SL{n+1}$ subgroup. Under this decomposition $\SL{n+1} \rightarrow \SL{n}$ with $V^I = \left( V^i,\, V^0 \right)$, where $i = 1, \ldots, n$, we solve the section condition by having physical coordinates $y^{0i}$ on the $n$-dimensional manifold $M$, \textit{i.e.} $\partial_{ij} = 0$ for all fields with only $\partial_{0i} = \partial_i \neq 0$. Correspondingly, the generalised tangent bundle, whose sections are in the antisymmetric representation of $\SL{n+1}$ and carry weight $\frac{n-3}{n+1}$, decomposes as
\begin{equation} \label{eq:SLnE}
	E = TM \oplus \Lambda^{n-2} T^*M \,,
\end{equation}
Note that for the case of interest to us, $n=3$, \eqref{eq:SLnE} reduces to
\begin{equation}
	E = TM \oplus T^*M \,,
\end{equation}
and its fibres transform in the $\mathbf{6}$ of $\SL{4} \simeq \SO{3,3}$. It is convenient to also introduce the generalised bundle with fibres in the anti-fundamental of $\SL{n+1}$, which is, similarly, given by
\begin{equation}
	N = T^*M \oplus \Lambda^{n} T^*M \,.
\end{equation}
Sections of $N$ are generalised tensors transforming in the anti-fundamental of $\SL{n+1}$ and carrying weight $\frac{2}{n+1}$. For $V, V' \in \Gamma(E)$ and $W \in \Gamma(N)$, the generalised Lie derivative \eqref{eq:SLnGenLie} reduces to
\begin{equation}
	\begin{split}
		\gL_{V} V' &= [v, v'] + L_{v} \omega'_{(n-2)} - \imath_{v'} d\omega_{(n-2)} \,, \\
		\gL_{V} W &= L_v \omega_{(1)} + L_v \omega_{(n)} + \omega_{(1)} \wedge d\omega_{(n-2)} \,,
	\end{split}
\end{equation}
where we write $V = v + \omega_{(n-2)}$, $V' = v' + \omega'_{(n-2)}$ and $W = \omega_{(1)} + \omega_{(n)}$ as a formal of vectors and $p$-forms, and $L$ denotes the ordinary Lie derivative and $[v,v']$ the ordinary Lie bracket between vector fields $v$ and $v'$.

We can now describe the $S^n$ consistent truncation using the $\SL{n+1}$ ExFT above. Let $Y^I$, $I = 1, \ldots, n+1$ be the embedding coordinates of $S^{n} \subset \mathbb{R}^{n+1}$, such that $Y^I\, Y_I = 1$. On $S^n$, we can define a parallelisation of $N$ using the generalised frame
\begin{equation}\label{eq:SnFrame}
	\cU^{\fl{I}} = dY^{\fl{I}} + Y^{\fl{I}}\, vol_{S^n} - dY^{\fl{I}} \wedge A \,,
\end{equation}
where $A$ is an $(n-1)$-potential with field strength $dA = (n-1) vol_{S^n}$, and
\begin{equation}
	vol_{S^n} = \frac{1}{n!}\epsilon_{I_1 I_2 \ldots I_{n+1}} Y^{I_1}\, dY^{I_2} \wedge \ldots \wedge dY^{I_{n+1}} \,,
\end{equation}
is the volume form on $S^n$. The frame in \eqref{eq:SnFrame} has the important property that the $n+1$ generalised tensors are nowhere vanishing since $dY^I = 0$ only when $Y^I = 1$. Therefore, \eqref{eq:SnFrame} provides a generalised parallelisation of $S^n$ in the $\SL{n+1}$ ExFT.

In a local basis, \eqref{eq:SnFrame} gives us a $\GL{n+1}$ matrix, $\cU_I{}^{\fl{I}}$, whose determinant allows us to define the scalar density
\begin{equation} \label{eq:rho}
	\rho = \left(\det \cU_I{}^{\fl{I}}\right)^{-1/2} = \left(\det \mathring{g}\right)^{-1/2} \,,
\end{equation}
with $\mathring{g}$ the round metric on $S^n$. Using \eqref{eq:SnFrame} and \eqref{eq:rho} we can define the $\SL{n}$ twist matrix
\begin{equation} \label{eq:SLnFrame}
	U_I{}^{\fl{I}} = \rho^{2/(n+1)}\, \cU_I{}^{\fl{I}} \,,
\end{equation}
and hence a generalised frame for $E$ or a generalised bundle in any other rep of $\SL{n+1}$. For example, in the generalised tangent bundle \eqref{eq:SLnE}, we have the generalised frame $\cU_{\fl{I}\fl{J}}{}^{IJ} = \left(\det \mathring{g}\right)^{1/2} \cU_{[\fl{I}}{}^{I} \cU_{\fl{J}]}{}^J$, with components given by
\begin{equation} \label{eq:SLnGenFrame}
	2\, \cU_{\fl{I}\fl{J}} = v_{\fl{I}\fl{J}} + \star dY_{\fl{I}\fl{J}} + \imath_{v_{\fl{I}\fl{J}}} A \,,
\end{equation}
where $v_{\fl{I}\fl{J}}$ are the $\SO{n+1}$ Killing vectors of the round $S^n$, matching precisely the $\SL{n+1}$ parallelisations constructed in \cite{Lee:2014mla}.

Using \eqref{eq:SLnGenLie}, we can easily check that the generalised frame  \eqref{eq:SLnGenFrame}, \eqref{eq:SLnTwist} defines a generalised Leibniz parallelisation. In particular, using $dA = (n-1) vol_{S^n}$, we obtain the embedding tensor
\begin{equation}
	\gL_{\cU_{\fl{I}\fl{J}}} \cU^{\fl{K}} = - X_{\fl{I}\fl{J}\fl{L}}{}^{\fl{K}}\, \cU^{\fl{L}} \,,
\end{equation}
with
\begin{equation} \label{eq:SLnEmbeddingTensor}
	X_{\fl{I}\fl{J},\fl{L}}{}^{\fl{K}} = - \delta_{\fl{L}[\fl{I}} \delta_{\fl{J}]}^{\fl{K}} \,.
\end{equation}

Finally, we can also connect the above with the $\SL{n+1}$ Ansatz construction of \cite{Hohm:2014qga}. We do this by evaluating \eqref{eq:rho} and \eqref{eq:SLnFrame} on the northern hemisphere $Y^I = \left( y^i,\, \sqrt{1-y^i y_i}\right)$, $i = 1, \ldots, n$ and using the gauge choice for the potential $A$
\begin{equation}
	A_{i_1 \ldots i_{n-1}} = \epsilon_{i_1 \ldots i_{n-1}j}\, y^j (1 + K(v)) \,,
\end{equation}
with $v = y^i\, y^i$ and $\epsilon_{i_1 \ldots i_{n-1}j}$ the volume form on $S^n$. With the above choices, we precisely recover the $\SL{n+1}$ twist matrix of \cite{Hohm:2014qga}, \textit{i.e.}
\begin{equation} \label{eq:SLnTwist}
	\UI_{\fl{I}}{}^I = \begin{pmatrix}
		\left(1-v\right)^{-1/(n+1)} \left( \delta_i{}^j + y_i\, y^j K(v) \right) & y_i \left(1-v\right)^{(n-1)/(2(n+1))} \\
		y^j \left(1-v\right)^{(n-1)/(2(n+1))} K(v) & \left(1-v\right)^{\tfrac{n}{n+1}}
	\end{pmatrix} \,.
\end{equation}

\section{Representations of \texorpdfstring{$D^{1}(2,1;\alpha)$}{D(2,1,a)}} \label{app:d21a}
The supergroup $D^{1}(2,1;\alpha)$ has
\begin{equation}
	{\rm SU}(1,1)\times{\rm SU}(2)\times\ts{{\rm SU}(2)}
\end{equation}
as its even part. Its multiplets are labelled by the spins $\ell$ and $\ts{\ell}$ of each ${\rm SU}(2)$ factor and the eigenvalue $h$ of the $\mathfrak{su}(1,1)\simeq\mathfrak{sl}(2,\mathbb{R})$ generator. Generic short multiplets $[\ell,\ts{\ell}]_{\rm s}$ are given by~\cite{deBoer:1999gea,Eberhardt:2017fsi}
\begin{equation}
	\begin{array}{cc}
	 h_{0} & \big(\ell,\ts{\ell}\big) \\[5pt]
	 h_{0} + \nicefrac{1}{2} & \big(\ell+\nicefrac{1}{2},\ts{\ell}-\nicefrac{1}{2}\big) \oplus \big(\ell-\nicefrac{1}{2},\ts{\ell}-\nicefrac{1}{2}\big) \oplus \big(\ell-\nicefrac{1}{2},\ts{\ell}+\nicefrac{1}{2}\big) \\[5pt]
	 h_{0}+1 & \big(\ell,\ts{\ell}-1\big) \oplus \big(\ell,\ts{\ell}\big) \oplus \big(\ell-1,\ts{\ell}\big) \\[5pt]
	 h_{0} + \nicefrac{3}{2} & \big(\ell-\nicefrac{1}{2},\ts{\ell}-\nicefrac{1}{2}\big)\,,
	\end{array}
\end{equation}
with $h_{0}=(\alpha\,\ell+\ts{\ell})/(1+\alpha)$. Shortenings occur for $\ell<1$ or $\ts{\ell}<1$~\cite{deBoer:1999gea}. For instance:
\begin{equation}
	[0,0]_{\rm s}:\quad h_{0}=0 \quad (0,0),\qquad  [\nicefrac{1}{2},\nicefrac{1}{2}]_{\rm s}: \begin{array}{cc}
	 h_{0}=\nicefrac{1}{2} & \big(\nicefrac{1}{2},\nicefrac{1}{2}\big) \\[5pt]
	 1 & \big(1,0\big) \oplus \big(0,0\big) \oplus \big(0,1\big) \\[5pt]
	 \nicefrac{3}{2} & \big(\nicefrac{1}{2},\nicefrac{1}{2}\big) \\[5pt]
	 2 & \big(0,0\big)\,.
	\end{array}
\end{equation}
Two short multiplets can combine into a long multiplets $[\ell,\ts{\ell}]$ as follows:
\begin{equation}
 	[\ell,\ts{\ell}] = [\ell,\ts{\ell}]_{\rm s} \oplus [\ell+\nicefrac{1}{2},\ts{\ell}+\nicefrac{1}{2}]_{\rm s}\,.
\end{equation}
The value of $h$ is then not constrained. The explicit content of the long representation $[\ell,\ts{\ell}]$  is
\begin{equation}
	\begin{array}{cc}
	 h & \big(\ell,\ts{\ell}\big) \\[5pt]
	 h + \nicefrac{1}{2} & \big(\ell+\nicefrac{1}{2},\ts{\ell}+\nicefrac{1}{2}\big) \oplus\big(\ell+\nicefrac{1}{2},\ts{\ell}-\nicefrac{1}{2}\big) \oplus \big(\ell-\nicefrac{1}{2},\ts{\ell}+\nicefrac{1}{2}\big) \oplus \big(\ell-\nicefrac{1}{2},\ts{\ell}-\nicefrac{1}{2}\big) \\[5pt]
	 h+1 & \big(\ell+1,\ts{\ell}\big) \oplus \big(\ell,\ts{\ell}-1\big) \oplus \big(\ell,\ts{\ell}\big) \oplus \big(\ell,\ts{\ell}\big) \oplus \big(\ell-1,\ts{\ell}\big)\oplus \big(\ell,\ts{\ell}+1\big) \\[5pt]
	 h + \nicefrac{3}{2} & \big(\ell+\nicefrac{1}{2},\ts{\ell}+\nicefrac{1}{2}\big) \oplus\big(\ell+\nicefrac{1}{2},\ts{\ell}-\nicefrac{1}{2}\big) \oplus \big(\ell-\nicefrac{1}{2},\ts{\ell}+\nicefrac{1}{2}\big) \oplus \big(\ell-\nicefrac{1}{2},\ts{\ell}-\nicefrac{1}{2}\big) \\[5pt]
	 h + 2 & \big(\ell,\ts{\ell}\big)\,.
	\end{array}
\end{equation}
	
\bibliographystyle{JHEP}
\bibliography{NewBib}
	
\end{document}